\documentclass[12pt]{article}
\usepackage{epsfig}
\usepackage{psfrag}
\usepackage{latexsym}
\usepackage{indentfirst}
\usepackage{fancyhdr}
\usepackage{amssymb}
\usepackage{amsfonts}
\usepackage{cite}
\usepackage{bbold}
\usepackage{color}
\usepackage[footnotesize]{caption2}
\usepackage{graphicx}
\usepackage[center,footnotesize,hang]{subfigure}
\usepackage{url}

\def\be{\begin{equation}}
\def\ee{\end{equation}}
\def\bc{\begin{center}}
\def\ec{\end{center}}
\def\bea{\begin{eqnarray}}
\def\eea{\end{eqnarray}}

\catcode`@=11
\def\marginnote#1{}
\newcount\hour
\newcount\minute
\newtoks\amorpm
\hour=\time\divide\hour by60
\minute=\time{\multiply\hour by60 \global\advance\minute by-\hour}
\edef\standardtime{{\ifnum\hour<12 \global\amorpm={am}%
        \else\global\amorpm={pm}\advance\hour by-12 \fi
        \ifnum\hour=0 \hour=12 \fi
        \number\hour:\ifnum\minute<10 0\fi\number\minute\the\amorpm}}
\edef\militarytime{\number\hour:\ifnum\minute<10 0\fi\number\minute}
\def\draftlabel#1{{\@bsphack\if@filesw {\let\thepage\relax
   \xdef\@gtempa{\write\@auxout{\string
      \newlabel{#1}{{\@currentlabel}{\thepage}}}}}\@gtempa
   \if@nobreak \ifvmode\nobreak\fi\fi\fi\@esphack}
        \gdef\@eqnlabel{#1}}
\def\@eqnlabel{}
\def\@vacuum{}
\def\draftmarginnote#1{\marginpar{\raggedright\scriptsize\tt#1}}
\def\draft{\oddsidemargin 0.0truein
        \def\@oddfoot{\sl preliminary draft \hfil
        \rm\thepage\hfil\sl\today\quad\militarytime}
        \let\@evenfoot\@oddfoot \overfullrule 3pt
        \let\label=\draftlabel
        \let\marginnote=\draftmarginnote
   \def\@eqnnum{(\theequation)\rlap{\kern\marginparsep\tt\@eqnlabel}%
\global\let\@eqnlabel\@vacuum}  }
\catcode`@=12
\begin{document}
\begin{titlepage}
\vspace*{-1cm}
\phantom{hep-ph/***}


\vskip 2.5cm
\begin{center}
{\Large\bf Fermion Masses and Flavor Mixings in a Model with $S_4$
Flavor Symmetry }
\end{center}
\vskip 0.2  cm
\vskip 0.5  cm
\begin{center}
{\large Gui-Jun Ding}~\footnote{e-mail address: dinggj@ustc.edu.cn}
\\
\vskip .2cm {\it Department of Modern Physics,}
\\
{\it University of Science and Technology of China, Hefei, Anhui 230026,
China}
\\
\end{center}
\vskip 0.7cm
\begin{abstract}
\noindent

We present a supersymmetric model of quark and lepton based on
$S_4\times Z_3\times Z_4$ flavor symmetry. The $S_4$ symmetry is
broken down to Klein four and $Z_3$ subgroups in the neutrino and
the charged lepton sectors respectively. Tri-Bimaximal mixing and
the charged lepton mass hierarchies are reproduced simultaneously at
leading order. Moreover, a realistic pattern of quark masses and
mixing angles is generated with the exception of the mixing angle
between the first two generations, which requires a small accidental
enhancement. It is remarkable that the mass hierarchies are
controlled by the spontaneous breaking of flavor symmetry in our
model. The next to leading order contributions are studied, all the
fermion masses and mixing angles receive corrections of relative
order $\lambda^2_c$ with respect to the leading order results. The
phenomenological consequences of the model are analyzed, the
neutrino mass spectrum can be normal hierarchy or inverted
hierarchy, and the combined measurement of the $0\nu2\beta$ decay
effective mass $m_{\beta\beta}$ and the lightest neutrino mass can
distinguish the normal hierarchy from the inverted hierarchy.

\end{abstract}
\end{titlepage}
\setcounter{footnote}{1}
 \vskip2truecm
%
\section{Introduction}
Neutrino has provided us a good window to the new physics beyond the
Standard Model. Neutrino oscillation experiments have provided solid
evidence that neutrinos have small but non-zero masses. Global data
fit to the current neutrino oscillation data demonstrates that the
mixing pattern in the leptonic sector is so different from the one
in the quark sector. Two independent fits for the mixing angles and
the mass squared differences are listed in Table \ref{tab:data_fit}.

\begin{table}[ht]\centering
\begin{tabular}{|c|cc|cc|} \hline\hline
        & \multicolumn{2}{|c|}{Ref.~\cite{Schwetz:2008er}}
        & \multicolumn{2}{|c|}{Ref.~\cite{GonzalezGarcia:2007ib}}\\
parameter   & best fit$\pm 1\sigma$ & 3$\sigma$ interval   & best
fit$\pm 1\sigma$ & 3$\sigma$ interval   \\ \hline
        $\Delta m^2_{21}\: [10^{-5}{\rm eV^2}]$
        & $7.65^{+0.23}_{-0.20}$ & 7.05--8.34
    & $7.67^{+0.22}_{-0.21}$ & 7.07--8.34
    \\[1mm]
        $\Delta m^2_{31}\: [10^{-3}{\rm eV^2}]$
        & $\pm 2.40^{+0.12}_{-0.11}$ & \hspace{-9pt}$\pm$(2.07--2.75)
    &
    \begin{tabular}{c}
        $-2.39 \pm 0.12$ \\
        $+2.49 \pm 0.12$
    \end{tabular}
    & \hspace{-11pt}
    \begin{tabular}{c}
        $-$(2.02--2.79) \\
        $+$(2.13--2.88)
    \end{tabular}
    \\[4mm]
        $\sin^2\theta_{12}$
        & $0.304^{+0.022}_{-0.016}$ & 0.25--0.37
        & $0.321^{+0.023}_{-0.022}$ & 0.26--0.40
    \\[2mm]
        $\sin^2\theta_{23}$
        & $0.50^{+0.07}_{-0.06}$ & 0.36--0.67
    & $0.47^{+0.07}_{-0.06}$ & 0.33--0.64
    \\[2mm]
        $\sin^2\theta_{13}$
        & $0.01^{+0.016}_{-0.011}$ & $\leq$ 0.056
        & $0.003\pm 0.015$ & $\leq$ 0.049
    \\
        \hline\hline
\end{tabular}
\caption{\label{tab:data_fit}%
Three flavour neutrino oscillation parameters from
two global data fits~\cite{Schwetz:2008er, GonzalezGarcia:2007ib}.}
\end{table}
As is obvious, the current neutrino oscillation data is remarkably
compatible with the so called Tri-Bimaximal (TB) mixing pattern
\cite{TBmix}, which suggests the following mixing pattern
\begin{equation}
\label{1}\sin^2\theta_{12,TB}=\frac{1}{3},~~\sin^2\theta_{23,TB}=\frac{1}{2},~~\sin^2\theta_{13,TB}=0
\end{equation}
These values lie in the $1\sigma$ range of global data analysis
shown in Table \ref{tab:data_fit}\footnote{$\sin^2\theta_{12,TB}$ is
exactly within the $1\sigma$ range of the second global data fit,
whereas it slightly above the $1\sigma$ up limit of the first fit.}.
Correspondingly, the leptonic Pontecorvo-Maki-Nakagawa-Sakata (PMNS)
mixing matrix is given by
\begin{equation}
\label{2}U^{TB}_{PMNS}=U_{TB}\;{\rm diag}(1,{\rm
e}^{i\alpha_{21}/2},{\rm e^{i\alpha_{31}/2}})
\end{equation}
where $\alpha_{21}$ and $\alpha_{31}$ are the Majorana CP violating
phases, and $U_{TB}$ is given by
\begin{equation}
\label{3}U_{TB}=\left(\begin{array}{ccc}
\sqrt{\frac{2}{3}}&\frac{1}{\sqrt{3}}&0\\
-\frac{1}{\sqrt{6}}&\frac{1}{\sqrt{3}}&\frac{1}{\sqrt{2}}\\
-\frac{1}{\sqrt{6}}&\frac{1}{\sqrt{3}}&-\frac{1}{\sqrt{2}}
\end{array}\right)
\end{equation}
The mixing in the quark sector is described by the famous CKM matrix
\cite{Cabibbo:1963yz}, and there is large mass hierarchies within
the quarks and charged leptons sectors respectively \cite{pdg}. The
origin of the observed fermion mass hierarchies and flavor mixings
is a great puzzle in particle physics. Nowadays promising candidates
for understanding such issue are the models based on spontaneously
breaking flavor symmetry, various models based on discrete or
continuous flavor symmetry have been proposed so far
\cite{continuous,horizontal}. Recently it was found that flavor
symmetry based on discrete group is particularly suitable to
reproduce specific mixing pattern at leading order \cite{review}.
The $A_4$ models are especially attractive, it has received
considerable interest in the recent past
\cite{TBModel,Altarelli:2005yp,Altarelli:2005yx,Altarelli:2008bg,Branco:2009by,Altarelli:2009kr}.
So far various $A_4$ flavor models have been proposed, and their
phenomenological consequences were analyzed
\cite{Bertuzzo:2009im,Hagedorn:2009jy,AristizabalSierra:2009ex,Felipe:2009rr}.
These models assumed that $A_4$ symmetry is realized at a high
energy scale, the lepton fields transform nontrivially under the
symmetry group, and the flavor symmetry is spontaneously broken by a
set of flavons with the vacuum expectation value (VEV) along a
specific direction. The misalignment in the flavor space between the
charged lepton and the neutrino sectors results in the TB lepton
mixing.

If extend the $A_4$ symmetry to the quark sector, the quark mixing
matrix $V_{CKM}$ turns out to be unity matrix at leading order
\cite{Altarelli:2005yx}, However, the subleading contributions of
the higher dimensional operators are too small to provide large
enough deviations of $V_{CKM}$ from the identity matrix. The
possible ways of resolving this issue are to consider new sources of
symmetry breaking or enlarge the symmetry group. Two discrete groups
$T'$
\cite{Ding:2008rj,Carr:2007qw,Feruglio:2007uu,Chen:2007afa,Frampton:2007et,Aranda:2007dp,Frampton:2008bz,Eby:2008uc}
and $S_4$
\cite{Ma:2005pd,Bazzocchi:2008ej,s4,Bazzocchi:2009da,Ishimori:2008fi,Altarelli:2009gn}
are found to be promising, both groups have two dimensional
irreducible representation, which is very useful to describing the
quark sector. The $S_4$ symmetry is particularly interesting, $S_4$
as a horizontal symmetry group has been proposed long ago
\cite{Pakvasa:1978tx}, and some models with different purposes have
been built \cite{Hagedorn:2006ug}. Recently it was claimed to be
minimal flavor group capable of yielding the TB mixing without fine
tuning \cite{Lam:2008rs,Lam:2008sh,Lam:2009hn}. However, Grimus et
al. were against this point \cite{Grimus:2009pg}.

In this work, we build a SUSY model based on $S_4\times Z_3\times
Z_4$ flavor group, the neutrino mass is generated via the
conventional type I See-Saw mechanism \cite{seesaw}. Our model
naturally produces the TB mixing and the charged lepton mass
hierarchy at leading order. Furthermore, we extend the model to the
quark sector, the realistic patterns of quark masses and mixing
angles are generated. In our model the mass hierarchies are
controlled by the spontaneous breaking of the flavor symmetry
instead of the Froggatt-Nielsen (FN) mechanism
\cite{Froggatt:1978nt}.

This article is organized as follows. Section 2 is the group theory
of $S_4$ group, where the subgroup, the equivalent class, and the
representation of $S_4$ are presented. In section 3 we justify the
vacuum alignment of our model in the supersymmetric limit. In
section 4 we present our model in both the lepton and quark sectors,
its basic features and theoretical predictions are discussed. In
section 5 we analyze the phenomenological implications of the model
in detail, which include the mass spectrum, neutrinoless double beta
decay and the Majorana CP violating phases etc. The corrections
induced by the next to leading order terms are studied in section 6.
Finally we summarize our results in the conclusion section.

\section{The discrete group $S_4$}

$S_4$ is the permutation group of 4 objects. The group has 24
distinct elements, and it can be generated by two elements $S$ and
$T$ obeying the relations
\begin{equation}
\label{4} S^4=T^3=1,~~~ST^2S=T
\end{equation}
Without loss of generality, we could choose
\begin{equation}
\label{5}S=(1234),~~~~~T=(123)
\end{equation}
where the cycle (1234) denotes the permutation
$(1,2,3,4)\rightarrow(2,3,4,1)$, and (123) means
$(1,2,3,4)\rightarrow(2,3,1,4)$. The 24 elements belong to 5
conjugate classes and are generated from $S$ and $T$ as follows
\begin{eqnarray*}
&&{\cal C}_1:1\\
&&{\cal
C}_2:\;STS^2=(12),\;TSTS^2=(13),\;ST^2=(14),\;S^2TS=(23),\;TST=(24),\;T^2S=(34)\\
&&{\cal C}_3:\;TS^2T^2=(12)(34),\;S^2=(13)(24),\;T^2S^2T=(14)(23)\\
&&{\cal
C}_4:\;T=(123),\;T^2=(132),\;T^2S^2=(124),\;S^2T=(142),\;S^2TS^2=(134),\;STS=(143),\\
&&~~~~~S^2T^2=(234),\;TS^2=(243)\\
&&{\cal
C}_5:\;S=(1234),\;T^2ST=(1243),\;ST=(1324),\;\;TS=(1342),\;TST^2=(1423),\;S^3=(1432)\\
\end{eqnarray*}
The structure of the group $S_4$ is rather rich, it has thirty
proper subgroups of orders 1, 2, 3, 4, 6, 8, 12 or 24. Concretely, the
subgroups of $S_4$ are as follows
\begin{enumerate}
\item{The trivial group only consisting of the unit element.}
\item{Six two-element subgroups generated by a transposition of the form $\{1,(ij)\}$ with $i\neq j$}

$H^{(1)}_2=\{1,STS^2\}$, $H^{(2)}_2=\{1,TSTS^2\}$,
$H^{(3)}_2=\{1,ST^2\}$, $H^{(4)}_2=\{1,S^2TS\}$,
$H^{(5)}_2=\{1,TST\}$ and $H^{(6)}_2=\{1,T^2S\}$
\item{Three two-element subgroups generated by a double transition of the form $\{1,(ij)(kl)\}$ with $i\neq j\neq k\neq l$ }

$H^{(7)}_2=\{1,TS^2T^2\}$, $H^{(8)}_2=\{1,S^2\}$,
$H^{(9)}_2=\{1,T^2S^2T\}$
\item{Four subgroups of order three, which is spanned by a three-cycle}

$H^{(1)}_3=\{1,T,T^2\}$,$H^{(2)}_3=\{1,T^2S^2,S^2T\}$,$H^{(3)}_3=\{1,S^2TS^2,STS\}$,$H^{(4)}_3=\{1,S^2T^2,TS^2\}$
\item{The four-element subgroups generated by a four-cycle, they are of the form $\{1,g,g^2,g^3\}$ with $g$ any four-cycle}

$H^{(1)}_4=\{1,S,S^2,S^3\}$,
$H^{(2)}_4=\{1,TS,T^2ST,T^2S^2T\}$,$H^{(3)}_4=\{1,
TST^2,ST,TS^2T^2\}$
\item{The four-element subgroups generated by two disjoint transpositions, which is isomorphic to Klein four group}

$H^{(4)}_4=\{1,STS^2,T^2S,TS^2T^2\}$,
$H^{(5)}_4=\{1,TSTS^2,TST,S^2\}$,
$H^{(6)}_4=\{1,ST^2,S^2TS,T^2S^2T\}$
\item{The order four subgroup comprising of the identity and three double transitions, which is isomorphic to Klein four group}

$H^{(7)}_4=\{1,TS^2T^2,S^2,T^2S^2T\}$
\item{Four subgroups of order six, which is isomorphic to $S_3$. They are the permutation groups of any three of the four objects, leaving the fourth invariant}

$H^{(1)}_6=\{1, STS^2,TSTS^2,S^2TS,T,T^2\}$,
$H^{(2)}_6=\{1,STS^2,ST^2,TST,T^2S^2,S^2T\}$, \\
$H^{(3)}_6=\{1,TSTS^2,ST^2,T^2S,S^2TS^2,STS\}$,
$H^{(4)}_6=\{1,S^2TS,TST,T^2S,S^2T^2,TS^2\}$
\item{Three eight-element subgroups, which is isomorphic to $D_4$ }

$H^{(1)}_8=\{1,TSTS^2,TST,S,S^3,TS^2T^2,S^2,T^2S^2T\}$,\\
$H^{(2)}_8=\{1,STS^2,T^2S,ST,TST^2,TS^2T^2,S^2,T^2S^2T\}$\\
$H^{(3)}_8=\{1,ST^2,S^2TS,T^2ST,TS,TS^2T^2,S^2,T^2S^2T\}$
\item{The alternating group $A_4$}

$A_4=\{1,TS^2T^2,S^2,T^2S^2T,T,T^2,T^2S^2,S^2T,S^2TS^2,STS,S^2T^2,TS^2\}$
\item{The whole group}
\end{enumerate}
In particular, $H^{(7)}_4$ and $A_4$ are the invariant subgroups of
$S_4$. Since the number of the unequivalent irreducible
representation is equal to the number of class, the $S_4$ group has
five irreducible representations: $1_1$, $1_2$, 2, $3_1$ and $3_2$.
$1_1$ is the identity representation and $1_2$ is the antisymmetric
one. The Young diagram for the two dimensional representations is
self associated, and the Young diagrams corresponding to the three
dimensional representations $3_1$ and $3_2$ are associated Young
diagrams. For the same group element, the representation matrices of
$3_1$ and $3_2$ are exactly the same if the element is an even
permutation. Whereas the overall signs are opposite if the group
element is an odd permutation. It is notable that $S_4$ together
with $T'$ is the smallest group containing one, two and three
dimensional representations. The character table of $S_4$ group is
shown in Table \ref{tab:character}.

\begin{table}
\begin{center}
\begin{tabular}{|c|c|c|c|c|c|}\hline\hline
   &\multicolumn{5}{|c|}{classes}\\\cline{2-6}
   &${\cal C}_1$&${\cal C}_2$&${\cal C}_3$&${\cal C}_4$&${\cal
C}_5$\\\hline

$n_{{\cal C}_i}$&1&6&3&8&6\\\hline

$h_{{\cal C}_i}$&1&2&2&3&4\\\hline

$1_1$&1&1&1&1&1\\\hline

$1_2$&1&-1&1&1&-1\\\hline

2&2&0&2&-1&0\\\hline

$3_1$&3&1&-1&0&-1\\\hline

$3_2$&3&-1&-1&0&1\\\hline\hline
\end{tabular}
\caption{\label{tab:character}Character table of the $S_4$ group.
$n_{{\cal C}_i}$ denotes the number of the elements contained in the
class ${\cal C}_i$, and $h_{{\cal C}_i}$ is the order of the
elements of ${\cal C}_i$.}
\end{center}
\end{table}

From the character table of the $S_4$ group, we can
straightforwardly obtain the multiplication rules between the
various representations
\begin{eqnarray}
\nonumber&&1_i\otimes1_j=1_{((i+j)\;{\rm mod}\;
2)+1},~~~~1_i\otimes2=2,~~~~1_i\otimes3_j=3_{((i+j)\;{\rm mod}\;
2)+1}\\
\nonumber&&2\otimes2=1_1\oplus1_2\oplus2,~~~~2\otimes3_i=3_1\oplus3_2,~~~~3_i\otimes3_i=1_1\oplus2\oplus3_1\oplus3_2,\\
\label{6}&&3_1\otimes3_2=1_2\oplus2\oplus3_2\oplus3_2, ~~~{\rm
with}~ i,j=1,2
\end{eqnarray}
The explicit representation matrices of the generators $S$, $T$ and
other group elements for the five irreducible representations are
listed in Appendix A. From these representation matrices, one
can explicitly calculate the Clebsch-Gordan coefficients for the
decomposition of the product representations, and the same results
as those in Ref.\cite{s4} are obtained.

%
\section{\label{sec:alignment}Field content and the vacuum alignment}

The model is supersymmetric and based on the discrete symmetry
$S_4\times Z_3\times Z_4$. Supersymmetry is introduced in order to
simplify the discussion of the vacuum alignment. The $S_4$ component
controls the mixing angles, the auxiliary $Z_3$ symmetry guarantees
the misalignment in flavor space between the neutrino and the
charged lepton mass eigenstates, and the $Z_4$ component is crucial
to eliminating the unwanted couplings and reproducing the observed
mass hierarchy. The fields of the model and their classification
under the flavor symmetry are shown in Table \ref{tab:trans}, where
two Higgses doublets $h_{u,d}$ of the minimal supersymmetric
standard model are present. If the $S_4$ flavor symmetry is
preserved until the electroweak scale, then all the fermions would
be massless. Therefore $S_4$ symmetry should be broken by the
suitable flavon fields, which are standard model singlets. Another
critical issue of the flavor model building is the vacuum alignment,
a global continuous $U(1)_R$ symmetry is exploited to simplify the
vacuum alignment problem. This symmetry is broken to the discrete R
parity once we include the gaugino mass in the model. The matter
fields carry +1 R-charge, the Higgses and the flavon supermultiplets
have R-charge 0. The spontaneous breaking of $S_4$ symmetry can be
implemented by introducing a new set of multiplets, the driving
fields carrying 2 unit R-charge. Consequently the driving fields
enter linearly into the superpotential. The suitable driving fields
and their transformation properties are shown in Table
\ref{tab:driving}.  In the following, we will discuss the
minimization of the scalar potential in the supersymmetric limit. At
the leading order, the most general superpotential dependent on the
driving fields, which is invariant under the flavor symmetry group
$S_4\times Z_3\times Z_4$, is given by

\begin{table}
\begin{center}
\begin{tabular}{|c|c|c|c|c|c|c|c|c|c|c|c|c|c|c||c|c|c|c|c|c|}\hline\hline
& $\ell$  & $e^{c}$ & $\mu^{c}$ &  $\tau^{c}$ &$\nu^c$ & $Q_L$ &
$Q_3$ & $u^{c}$& $c^{c}$ & $t^{c}$ & $d^c$& $s^c$ &$b^{c}$ &
$h_{u,d}$&$\varphi$ & $\chi$ & $\theta$ & $\eta$ & $\phi$ & $\Delta$
\\\hline

$\rm{S_4}$& $3_1$& $1_1$ & $1_2$&$1_1$ &  $3_1$  & 2  & $1_1$ &
$1_1$ &$1_2$ & $1_1$&$1_1$&$1_2$&$1_2$ & $1_1$&$3_1$ & $3_2$& $1_2$
& 2   & $3_1$   & $1_2$  \\\hline

$\rm{Z_{3}}$& $\omega$ & $\omega^2$& $\omega^2$&  $\omega^2$  & $1$
& 1& 1& 1& 1&1 & $\omega$&1&1&1 & 1 &1 &1 &  $\omega^2$  &
$\omega^2$   & $\omega^2$
\\\hline

$\rm{Z_{4}}$& 1 &i & -1&  -i  & 1  & -1 & 1 & i & 1 &1 &1&1&1  & 1&
i &i & 1&  1  & 1  & -1  \\\hline\hline
\end{tabular}
\end{center}
\caption{\label{tab:trans}The transformation rules of the matter fields and the flavons
under the symmetry groups $S_4$, $Z_3$ and $Z_4$. $\omega$ is the
third root of unity, i.e.
$\omega=e^{i\frac{2\pi}{3}}=(-1+i\sqrt{3})/2$. We denote $Q_L=(Q_1,Q_2)^t$ which are doublets of $S_4$, where $Q_1=(u,d)^t$ and $Q_2=(c,s)^t$ are
the electroweak SU(2) doublets of the first two generations. $Q_3=(t,b)^t$ is the electroweak
SU(2) doublet of the third generation. }
\end{table}

\begin{table}[hptb]
\begin{center}
\begin{tabular}{|c|c|c|c|c|c|c|}\hline\hline
Fields& $\varphi^{0}$  & $\xi^{'0}$ & $~\theta^{0}~$ & $~\eta^{0}~$
& $~\phi^{0}~$ & $~\Delta^{0}~$ \\\hline

$\rm{S_4}$& $3_1$  & $1_2$& $1_1$ & 2   & $3_2$  & $1_1$
\\\hline

$\rm{Z_{3}}$& 1 &1 & 1& $\omega^2$   & $\omega^2$   & $\omega^2$
\\\hline

$\rm{Z_{4}}$& -1 &-1 &1 &  1  &  1 & 1 \\\hline\hline
\end{tabular}
\caption{\label{tab:driving} The driving fields and their
transformation properties under the flavor group $S_4\times Z_3\times
Z_4$. }
\end{center}
\end{table}
\begin{eqnarray}
\nonumber &&w_v=g_1(\varphi^{0}(\varphi\varphi)_{3_1})_{1_1}+g_2(\varphi^{0}(\chi\chi)_{3_1})_{1_1}+g_3(\varphi^{0}(\varphi\chi)_{3_1})_{1_1}+g_4\xi^{'0}(\varphi\chi)_{1_2}+M^2_{\theta}\theta^{0}+\kappa\theta^{0}\theta^2+\\
\label{7}&&f_1(\eta^0(\eta\eta)_2)_{1_1}+f_2(\eta^{0}(\phi\phi)_2)_{1_1}+f_3(\phi^{0}(\eta\phi)_{3_2})_{1_1}+h_1\Delta^{0}\Delta^2+h_2\Delta^{0}(\eta\eta)_{1_1}+h_3\Delta^{0}(\phi\phi)_{1_1}
\end{eqnarray}
where the subscript $1_1$ denotes the contraction in $1_1$, similar rule applies to other subscripts $1_2$, $2$, $3_1$ and $3_2$.
In the SUSY limit, the vacuum configuration is determined by the
vanishing of the derivative of $w_v$ with respect to each component
of the driving fields
\begin{eqnarray}
\nonumber&&\frac{\partial w_v}{\partial
\varphi^{0}_1}=2g_1(\varphi^2_1-\varphi_2\varphi_3)+2g_2(\chi^2_1-\chi_2\chi_3)+g_3(\varphi_2\chi_3-\varphi_3\chi_2)=0\\
\nonumber&&\frac{\partial
w_v}{\partial\varphi^{0}_2}=2g_1(\varphi^2_2-\varphi_1\varphi_3)+2g_2(\chi^2_2-\chi_1\chi_3)+g_3(\varphi_3\chi_1-\varphi_1\chi_3)=0\\
\nonumber&&\frac{\partial w_v}{\partial
\varphi^{0}_3}=2g_1(\varphi^2_3-\varphi_1\varphi_2)+2g_2(\chi^2_3-\chi_1\chi_2)+g_3(\varphi_1\chi_2-\varphi_2\chi_1)=0\\
\nonumber&&\frac{\partial w_v}{\partial
\xi^{'0}}=g_4(\varphi_1\chi_1+\varphi_2\chi_3+\varphi_3\chi_2)=0\\
\label{8}&&\frac{\partial w_v}{\partial
\theta^{0}}=M^2_{\theta}+\kappa\theta^2=0
\end{eqnarray}
This set of equations admit the solution
\begin{eqnarray}
\label{9}&&\langle\varphi\rangle=(0,v_{\varphi},0),~~~\langle\chi\rangle=(0,v_{\chi},0),~~~\langle\theta\rangle=v_{\theta}
\end{eqnarray}
with
\begin{equation}
\label{10}v^2_{\varphi}=-\frac{g_2}{g_1}v^2_{\chi},
~~~~v^2_{\theta}=-\frac{M^2_{\theta}}{\kappa},~~~v_{\chi} {\rm~
undetermined}
\end{equation}
From the driving superpotential $w_v$, we can also derive the
equations from which to extract the vacuum expectation values of
$\eta$, $\phi$ and $\Delta$
\begin{eqnarray}
\nonumber&&\frac{\partial w_v}{\partial
\eta^{0}_1}=f_1\eta^2_1+f_2(\phi^2_3+2\phi_1\phi_2)=0\\
\nonumber&&\frac{\partial w_v}{\partial
\eta^{0}_2}=f_1\eta^2_2+f_2(\phi^2_2+2\phi_1\phi_3)=0\\
\nonumber&&\frac{\partial w_v}{\partial
\phi^{0}_1}=f_3(\eta_1\phi_2-\eta_2\phi_3)=0\\
\nonumber&&\frac{\partial w_v}{\partial
\phi^{0}_2}=f_3(\eta_1\phi_1-\eta_2\phi_2)=0\\
\nonumber&&\frac{\partial w_v}{\partial
\phi^{0}_3}=f_3(\eta_1\phi_3-\eta_2\phi_1)=0\\
\label{11}&&\frac{\partial w_v}{\partial
\Delta^0}=h_1\Delta^2+2h_2\eta_1\eta_2+h_3(\phi^2_1+2\phi_2\phi_3)=0
\end{eqnarray}
The solution to the above six equations is
\begin{eqnarray}
\label{12}\langle\eta\rangle=(v_{\eta},v_{\eta}),~~~~\langle\phi\rangle=(v_{\phi},v_{\phi},v_{\phi}),~~~~\langle\Delta\rangle=v_{\Delta}
\end{eqnarray}
with the conditions
\begin{equation}
\label{13}v^2_{\phi}=-\frac{f_1}{3f_2}v^2_{\eta},~~~~v^2_{\Delta}=\frac{f_1h_3-2f_2h_2}{f_2h_1}v^2_{\eta},~~~~v_{\eta} {\rm~
undetermined}
\end{equation}
The vacuum expectation values (VEVs) of the flavons can be very
large, much larger than the electroweak scale, and we expect that
all the VEVs are of a common order of magnitude. This is a very
common assumption in the flavor model building, which guarantees the
reasonability of the subsequent perturbative expansion in inverse
power of the cutoff scale $\Lambda$. Acting on the vacuum
configurations of Eq.(\ref{9}) and Eq.(\ref{12}) with the elements
of the flavor symmetry group $S_4$, we can see that the VEVs of
$\eta$ and $\phi$ are invariant under four elements 1, $TST$,
$TSTS^2$ and $S^2$, which exactly constitute the Klein four group
$H^{(5)}_4$. On the contrary, the VEVs of $\varphi$ and $\chi$ break
$S_4$ completely. Under the action of $T$ or $T^2$, the directions
of $\langle\varphi\rangle$ and $\langle\chi\rangle$ are invariant
except an overall phase. Considering the enlarged group $S_4\times
Z_3$, the vacuum configuration Eq.(9) preserves the subgroup $Z_3$
generated by $\omega T$, which is defined as the simultaneous
transformation of $T\in S_4$ and $\omega\in Z_3$. As we shall see
later that the $S_4$ flavor symmetry is spontaneously broken down by
the VEVs of $\eta$ and $\phi$ in the neutrino sector at the leading
order(LO), and it is broken down by the VEVs of $\varphi$ and $\chi$
in the charged lepton sector. Whereas both $\eta$, $\phi$ and
$\varphi$, $\chi$ are involved in generating the quark masses. The
$S_4$ flavor symmetry is broken into the Klein four symmetry
$H^{(5)}_4$ and the $Z_3$ symmetry generated by $T$ in the neutrino
and the charged lepton sector respectively at LO. This symmetry
breaking chain is crucial to generating the TB mixing.


\section{The model with $S_4\times Z_3\times Z_4$ flavor symmetry}
In this section we shall propose a concise supersymmetric (SUSY)
model based on $S_4\times Z_3\times Z_4$ flavor symmetry with the
vacuum alignment of Eq.(9) and Eq.(12).
\subsection{Charged leptons}
The charged lepton masses are described by the following
superpotential
\begin{eqnarray}
\nonumber&&w_{\ell}=\frac{y_{e1}}{\Lambda^3}\;e^{c}(\ell\varphi)_{1_1}(\varphi\varphi)_{1_1}h_d+\frac{y_{e2}}{\Lambda^3}\;e^{c}((\ell\varphi)_2(\varphi\varphi)_2)_{1_1}h_d+\frac{y_{e3}}{\Lambda^3}\;e^{c}((\ell\varphi)_{3_1}(\varphi\varphi)_{3_1})_{1_1}h_d\\
\nonumber&&~~+\frac{y_{e4}}{\Lambda^3}\;e^{c}((\ell\chi)_2(\chi\chi)_2)_{1_1}h_d+\frac{y_{e5}}{\Lambda^3}\;e^{c}((\ell\chi)_{3_1}(\chi\chi)_{3_1})_{1_1}h_d+\frac{y_{e6}}{\Lambda^3}\;e^{c}(\ell\varphi)_{1_1}(\chi\chi)_{1_1}h_d\\
\nonumber&&~~+\frac{y_{e7}}{\Lambda^3}\;e^{c}((\ell\varphi)_2(\chi\chi)_2)_{1_1}h_d+\frac{y_{e8}}{\Lambda^3}\;e^{c}((\ell\varphi)_{3_1}(\chi\chi)_{3_1})_{1_1}h_d+\frac{y_{e9}}{\Lambda^3}\;e^{c}((\ell\chi)_2(\varphi\varphi)_2)_{1_1}h_d\\
\label{14}&&~~+\frac{y_{e10}}{\Lambda^3}\;e^{c}((\ell\chi)_{3_1}(\varphi\varphi)_{3_1})_{1_1}h_d+\frac{y'_{\mu}}{\Lambda^2}\mu^{c}(\ell(\varphi\chi)_{3_2})_{1_2}h_d+\frac{y_{\tau}}{\Lambda}\tau^{c}(\ell\varphi)_{1_1}h_d+...
\end{eqnarray}
In the above superpotential $w_{\ell}$, for each charged lepton,
only the lowest order operators in the expansion in powers of
$1/\Lambda$ are displayed explicitly. Dots stand for higher
dimensional operators. Note that the auxiliary $Z_4$ symmetry
imposes different powers of $\varphi$ and $\chi$ for the electron,
mu and tau terms. At LO only the tau mass is generated, the muon and
the electron masses are generated by high order contributions. After
the flavor symmetry breaking and the electroweak symmetry breaking,
the charged leptons acquire masses, and $w_{\ell}$ becomes
\begin{eqnarray}
\nonumber&&w_{\ell}=\Big[(y_{e2}-2y_{e3})\frac{v^3_{\varphi}}{\Lambda^3}+(-y_{e4}+2y_{e5})\frac{v^3_{\chi}}{\Lambda^3}+(y_{e7}-2y_{e8})\frac{v_{\varphi}v^2_{\chi}}{\Lambda^3}+(-y_{e9}+2y_{e10})\frac{v_{\chi}v^2_{\varphi}}{\Lambda^3}\Big]v_de^{c}e\\
\nonumber&&~~~~+2y'_{\mu}\frac{v_{\varphi}v_{\chi}}{\Lambda^2}\;v_{d}\mu^{c}\mu+y_{\tau}\frac{v_{\varphi}}{\Lambda}\;v_d\tau^{c}\tau\\
\label{15}&&~~\equiv
y_e\frac{v^3_{\varphi}}{\Lambda^3}\;v_de^{c}e+y_{\mu}\frac{v_{\varphi}v_{\chi}}{\Lambda^2}\;v_{d}\mu^{c}\mu+y_{\tau}\frac{v_{\varphi}}{\Lambda}\;v_d\tau^{c}\tau
\end{eqnarray}
where $v_d=\langle h_d\rangle$,
$y_e=y_{e2}-2y_{e3}+(-y_{e4}+2y_{e5})\frac{v^3_{\chi}}{v^3_{\varphi}}+(y_{e7}-2y_{e8})\frac{v^2_{\chi}}{v^2_{\varphi}}+(-y_{e9}+2y_{e10})\frac{v_{\chi}}{v_{\varphi}}$
and $y_{\mu}=2y'_{\mu}$. As a result, the charged lepton mass matrix
is diagonal at LO
\begin{equation}
\label{16}m_{\ell}=\left(\begin{array}{ccc}
y_e\frac{v^3_{\varphi}}{\Lambda^3}&0&0\\
0&y_{\mu}\frac{v_{\varphi}v_{\chi}}{\Lambda^2}&0\\
0&0&y_{\tau}\frac{v_{\varphi}}{\Lambda}
\end{array}\right)v_d
\end{equation}
It is obvious that the hermitian matrix $m^{\dagger}_{\ell}m_{\ell}$ is
invariant under both $T$ and $T^2$ displayed in the Appendix A, i.e.,
\begin{equation}
\label{17}T^{\dagger}m^{\dagger}_{\ell}m_{\ell}T=m^{\dagger}_{\ell}m_{\ell}
\end{equation}
Conversely, the general matrix invariant under $T$ and $T^{2}$ must
be diagonal. Consequently the $S_4$ symmetry is broken to the $Z_3$
subgroup $H^{(1)}_1\equiv G_{\ell}$ in the charged lepton sector. The charged lepton masses can
be read out directly as
\begin{equation}
\label{18}m_e=\Big|y_e\frac{v^3_{\varphi}}{\Lambda^3}v_d\Big|,~~~m_{\mu}=\Big|y_{\mu}\frac{v_{\varphi}v_{\chi}}{\Lambda^2}v_d\Big|,~~~m_{\tau}=\Big|y_{\tau}\frac{v_{\varphi}}{\Lambda}v_d\Big|
\end{equation}
we notice that the charged lepton mass hierarchies are naturally
generated by the spontaneous symmetry breaking of $S_4$ symmetry
without exploiting the FN mechanism \cite{Altarelli:2005yx}. Using
the experimental data on the ratio of the lepton masses, one can
estimate the order of magnitude of $v_{\varphi}/\Lambda$ and
$v_{\chi}/\Lambda$. Assuming that the coefficients $y_e$, $y_{\mu}$
and $y_{\tau}$ are of ${\cal O}(1)$, we obtain
\begin{eqnarray}
\nonumber&&\frac{m_e}{m_{\tau}}\sim\frac{v^2_{\varphi}}{\Lambda^2}\simeq3\times10^{-4}\\
\label{19}&&\frac{m_{\mu}}{m_{\tau}}\sim\frac{v_{\chi}}{\Lambda}\simeq6\times10^{-2}
\end{eqnarray}
Obviously the solution to the above equations is
\begin{equation}
\label{20}(\frac{v_{\varphi}}{\Lambda},~\frac{v_{\chi}}{\Lambda})\sim(\pm1.73\times10^{-2},~6\times10^{-2})
\end{equation}
we see that the amplitudes of both $v_{\varphi}/\Lambda$ and
$v_{\chi}/\Lambda$ are roughly of the same order about ${\cal
O}(\lambda^2_c)$, where $\lambda_c$ is the Cabibbo angle.

\subsection{Neutrinos}
The superpotential contributing to the neutrino mass is as follows
\begin{eqnarray}
\label{21}w_{\nu}=\frac{y_{\nu1}}{\Lambda}((\nu^{c}\ell)_2\eta)_{1_1}h_u+\frac{y_{\nu2}}{\Lambda}((\nu^{c}\ell)_{3_1}\phi)_{1_1}h_u+\frac{1}{2}M(\nu^c\nu^c)_{1_1}+...
\end{eqnarray}
where dots denote the higher order contributions, $M$ is a constant
with dimension of mass, and the factor $\frac{1}{2}$ is a
normalization factor for convenience. The first two terms in
Eq.(\ref{21}) determine the neutrino Dirac mass matrix, and the
third term is Majorana mass term. After electroweak and $S_4$
symmetry breaking, we obtain the following LO contributions to the
neutrino Dirac and Majorana mass matrices
\begin{equation}
\label{22}m^{D}_{\nu}=\left(\begin{array}{ccc}2b&a-b&a-b\\
a-b&a+2b&-b\\
a-b&-b&a+2b\end{array}\right)v_{u},~~~~M_N=\left(\begin{array}{ccc}
M&0&0\\
0&0&M\\
0&M&0\end{array}\right)
\end{equation}
where $v_{u}=\langle h_u\rangle$,
$a=y_{\nu1}\frac{v_{\eta}}{\Lambda}$ and
$b=y_{\nu2}\frac{v_{\phi}}{\Lambda}$. We notice that the Dirac mass
matrix is symmetric and it is controlled by two parameters $a$ and
$b$. The eigenvalues of the Majorana matrix $M_N$ are given by
\begin{equation}
\label{23}M_1=M,~~M_2=M,~~M_3=-M
\end{equation}
The right handed neutrino masses are exactly degenerate, this is a
remarkable feature of our model. Integrating out the heavy degrees
of freedom, we get the light neutrino mass matrix, which is given by
the famous See-Saw relation
\begin{equation}
\label{24}m_{\nu}=-(m^{D}_{\nu})^{T}M^{-1}_{N}m^{D}_{\nu}=-\frac{v^2_u}{M}\left(\begin{array}{ccc}2a^2+6b^2-4ab&a^2-3b^2+2ab&a^2-3b^2+2ab\\
a^2-3b^2+2ab&a^2-3b^2-4ab&2a^2+6b^2+2ab\\
a^2-3b^2+2ab&2a^2+6b^2+2ab&a^2-3b^2-4ab
\end{array}\right)
\end{equation}
The above light neutrino mass matrix $m_{\nu}$ is
$2\leftrightarrow3$ invariant and it satisfies the magic symmetry
$(m_{\nu})_{11}+(m_{\nu})_{13}=(m_{\nu})_{22}+(m_{\nu})_{23}$.
Therefore it is exactly diagonalized by the TB mixing
\begin{equation}
\label{25}U^{T}_{\nu}m_{\nu}U_{\nu}={\rm diag}(m_1,m_2,m_3)
\end{equation}
The unitary matrix $U_{\nu}$ is written as
\begin{equation}
\label{26}U_{\nu}=U_{TB}\,{\rm
diag}(e^{-i\alpha_1/2},e^{-i\alpha_2/2},e^{-i\alpha_3/2})
\end{equation}
The phases $\alpha_{1}$, $\alpha_2$ and $\alpha_3$ are given by
\begin{eqnarray}
\nonumber&&\alpha_1={\rm arg}(-(a-3b)^2/M)\\
\nonumber&&\alpha_2={\rm arg}(-4a^2/M)\\
\label{27}&&\alpha_3={\rm arg}((a+3b)^2/M)
\end{eqnarray}
$m_{1}$, $m_{2}$ and $m_{3}$ in Eq.(2\ref5{}) are the light neutrino masses,
\begin{eqnarray}
\nonumber&&m_1=|(a-3b)^2|\frac{v^2_u}{|M|}\\
\nonumber&&m_2=4|a^2|\frac{v^2_u}{|M|}\\
\label{28}&&m_3=|(a+3b)^2|\frac{v^2_u}{|M|}
\end{eqnarray}
Concerning the neutrinos, the $S_4$ symmetry is spontaneously broken
by the VEVs of $\eta$ and $\phi$ at the LO. since both
$\langle\eta\rangle$ and $\langle\phi\rangle$ are invariant under
the actions of $TSTS^2$, $TST$ and $S^2$, the flavor symmetry $S_4$
is broken down to the Klein four subgroup $G_{\nu}\equiv
H^{(5)}_4=\{1,TSTS^2,TST,S^2\}$ in the neutrino sector. We can
straightforwardly check that the light neutrino mass matrix
$m_{\nu}$ is really invariant under $TSTS^2$, $TST$ and $S^2$. On
the contrary, the most general neutrino mass matrix invariant under
the Klein four group $G_{\nu}$ is given by
\begin{equation}
\label{29}m_{\nu}=\left(\begin{array}{ccc}
m_{11}&m_{12}&m_{12}\\
m_{12}&m_{22}&m_{11}+m_{12}-m_{22}\\
m_{12}&m_{11}+m_{12}-m_{22}& m_{22}
\end{array}\right)
\end{equation}
where $m_{11}$, $m_{12}$ and $m_{22}$ are arbitrary parameters. In
the present model, the light neutrino mass matrix is given by
Eq.(\ref{24}), which is a particular version of the neutrino mass
matrix in Eq.(\ref{29}). Since only two parameters $a$ and $b$ are
involved in our model, additional constraint has to be satisfied,
i.e. $3m^2_{11}+4m_{12}m_{11}-4m_{22}m_{11}-8m^2_{12}-4m^2_{22}=0$,
which is generally not implied by the invariance under $G_{\nu}$.
This is because that in our model the fields which break $S_4$ are a
doublet $\eta$ and a triplet $\phi$, there are no further flavons
transforming as $1_1$ or $3_2$ which couple to the neutrino sector.

In short summary, at the LO the $S_4$ flavor symmetry is broken down
to $Z_3$ and Klein four subgroup in the charged lepton and neutrino
sector respectively. We have obtained a diagonal and hierarchical
charged lepton mass matrix, the heavy neutrino masses are
degenerate, and the neutrino mixing matrix is exactly the TB matrix.

\subsection{Effective operators}





In the previous section, the neutrinos acquire masses via the
See-Saw mechanism. It is interesting to note that higher dimension
Weinberg operator cloud also contribute to the neutrino mass
directly, which may correspond to exchanging some heavy particles
rather than the right handed neutrinos $\nu^{c}$. In the present
model, these effective light neutrino mass operators are
\begin{eqnarray}
\nonumber&&w^{eff}_{\nu}=\frac{x}{\Lambda^3}(\ell h_u\ell
h_u)_{1_1}\Delta^2+\frac{y_1}{\Lambda^3}(\ell h_u\ell
h_u)_{1_1}(\eta^2)_{1_1}+\frac{y_2}{\Lambda^3}((\ell h_u\ell
h_u)_2(\eta^2)_2)_{1_1}+\frac{z_1}{\Lambda^3}(\ell h_u\ell
h_u)_{1_1}(\phi^2)_{1_1}\\
\label{30}~~~~~~&&+\frac{z_2}{\Lambda^3}((\ell h_u\ell
h_u)_2(\phi^2)_2)_{1_1}+\frac{z_3}{\Lambda^3}((\ell h_u\ell
h_u)_{3_1}(\phi^2)_{3_1})_{1_1}+\frac{w}{\Lambda^3}((\ell h_u\ell
h_u)_{3_1}(\eta\phi)_{3_1})_{1_1}
\end{eqnarray}
With the vacuum configurations displayed in Eq.(\ref{12}), the high
dimension operators $w^{eff}_{\nu}$ leads to the following effective
light neutrino mass matrix
\begin{equation}
\label{31}m^{eff}_{\nu}=\left(\begin{array}{ccc}
\alpha+2\gamma&\beta-\gamma&\beta-\gamma\\
\beta-\gamma&\beta+2\gamma&\alpha-\gamma\\
\beta-\gamma&\alpha-\gamma&\beta+2\gamma\end{array}\right)\frac{v^2_u}{\Lambda}
\end{equation}
where
\begin{eqnarray}
\nonumber&&\alpha=2x\frac{v^2_{\Delta}}{\Lambda^2}+4y_1\frac{v^2_{\eta}}{\Lambda^2}+6z_1\frac{v^2_{\phi}}{\Lambda^2}\\
\nonumber&&\beta=2y_2\frac{v^2_{\eta}}{\Lambda^2}+6z_2\frac{v^2_{\phi}}{\Lambda^2}\\
\label{32}&&\gamma=4w\frac{v_{\eta}v_{\phi}}{\Lambda^2}
\end{eqnarray}
Obviously $m^{eff}_{\nu}$ have the same texture as that in Eq.(29),
and it is remarkable that this mass matrix is diagonalized by TB
matrix,
\begin{equation}
\label{33}U^{T}_{TB}m^{eff}_{\nu}U_{TB}={\rm
diag}(m^{eff}_1,m^{eff}_2,m^{eff}_3)
\end{equation}
where $m^{eff}_1$, $m^{eff}_2$ and $m^{eff}_3$ are the effective
light neutrino masses coming from the above high dimension Weinberg
operators, they are given by
\begin{eqnarray}
\nonumber&&m^{eff}_1=(\alpha-\beta+3\gamma)\frac{v^2_u}{\Lambda}\\
\nonumber&&m^{eff}_1=(\alpha+2\beta)\frac{v^2_u}{\Lambda}\\
\label{34}&&m^{eff}_3=(-\alpha+\beta+3\gamma)\frac{v^2_u}{\Lambda}
\end{eqnarray}
If we consider the parameters $y_{\nu1,\nu2}\sim{\cal O}(1)$,
$y_{1,2}\sim{\cal O}(1)$, $z_{1,2,3}\sim{\cal O}(1)$ and $x\sim
w\sim{\cal O}(1)$, we get the ratio
\begin{equation}
\label{35}\frac{m^{eff}_i}{m_i}\sim\frac{M}{\Lambda}
\end{equation}
We see that the importance of Weinberg operators depends on the
relative size of $M$ and $\Lambda$. Since we have assumed that the
light neutrino masses mainly come from the See-Saw mechanism, the
right handed neutrino mass $M$ should be much smaller than the
cutoff scale $\Lambda$. In the context of a grand unified theory,
this corresponds to the requirement that $M$ is of order ${\cal
O}(M_{GUT})$ rather than of ${\cal O}(M_{Planck})$. In some flavor
models, right handed neutrino masses are required to be below the
cutoff $\Lambda$ as well, in order to reproduce the experimental
value of the small parameter $\Delta m^2_{sol}/\Delta m^2_{atm}$
\cite{Altarelli:2009kr,Altarelli:2009gn}. Another convenient way of
suppressing the contributions of the effective operators is to
introduce auxiliary symmetry further, so that the Weinberg operators
arise at much higher order and its contributions can be neglected.
\subsection{Extension to the quark sector}
The Yukawa superpotentials in the quark sector are
\begin{equation}
\label{36}w_{q}=w_u+w_d
\end{equation}
In the up quark sector, we have
\begin{eqnarray}
\nonumber&&w_u=y_tt^{c}Q_3h_u+\sum^{3}_{i=1}\frac{y_{ti}}{\Lambda^2}t^{c}(Q_L{\cal
O}^{(1)}_i)_{1_1}h_u+\sum^{2}_{i=1}\frac{y'_{ti}}{\Lambda^3}t^{c}(Q_L{\cal
O}^{(2)}_i)_{1_1}h_u+\sum^{3}_{i=1}\frac{y_{ci}}{\Lambda^2}c^{c}(Q_L{\cal
O}^{(1)}_i)_{1_2}h_u\\
\nonumber&&~~~~+\sum^{2}_{i=1}\frac{y'_{ci}}{\Lambda^3}c^{c}(Q_L{\cal
O}^{(2)}_i)_{1_2}h_u+\frac{y_{ct}}{\Lambda}c^cQ_3\theta
h_u+\sum^{8}_{i=1}\frac{y_{ui}}{\Lambda^4}u^{c}(Q_L{\cal
O}^{(3)}_i)_{1_1}h_u\\
\label{37}&&~~~~+\sum^{2}_{i=1}\frac{y'_{ui}}{\Lambda^3}u^{c}Q_3({\cal
O}^{(4)})_{1_1}h_u+...
\end{eqnarray}
where
\begin{eqnarray}
\nonumber&&{\cal O}^{(1)}=\{\varphi\varphi, \varphi\chi, \chi\chi\}\\
\nonumber&&{\cal O}^{(2)}=\{\eta^2\Delta, \phi^2\Delta\}\\
\nonumber&&{\cal
O}^{(3)}=\{\varphi\phi^3,\chi\phi^3,\varphi\eta\phi^2,\chi\eta\phi^2,\varphi\eta^2\phi,\chi\eta^2\phi,\varphi\phi\Delta^2,\chi\phi\Delta^2\}\\
\label{38}&&{\cal O}^{(4)}=\{\varphi^3,\varphi\chi^2\}
\end{eqnarray}

The superpotentials contributing to the down quark masses are as
follows
\begin{eqnarray}
\nonumber&&w_d=\frac{y_b}{\Lambda}b^cQ_3\theta
h_d+\sum^{3}_{i=1}\frac{y_{bi}}{\Lambda^2}b^{c}(Q_L{\cal
O}^{(1)}_i)_{1_2}h_d+\sum^{2}_{i=1}\frac{y'_{bi}}{\Lambda^3}b^c(Q_L{\cal
O}^{(2)}_i)_{1_2}h_d+\sum^{3}_{i=1}\frac{y_{si}}{\Lambda^2}s^{c}(Q_L{\cal
O}^{(1)}_i)_{1_2}h_d\\
\nonumber&&~~~~+\sum^{2}_{i=1}\frac{y'_{si}}{\Lambda^3}s^{c}(Q_L{\cal
O}^{(2)}_i)_{1_2}h_d+\sum^{3}_{i=1}\frac{y''_{si}}{\Lambda^3}s^{c}Q_3({\cal
O}^{(5)}_i)_{1_2}h_d+\sum^{6}_{i=1}\frac{y_{di}}{\Lambda^3}d^{c}(Q_L{\cal
O}^{(6)}_i)_{1_1}h_d\\
\label{39}&&~~~~+\frac{y'_{d1}}{\Lambda^3}d^{c}Q_3(\varphi\chi)_{1_2}\Delta
h_d+\sum^{9}_{i=1}\frac{y''_{di}}{\Lambda^4}d^{c}Q_3({\cal
O}^{(7)}_i)_{1_1}h_d+...
\end{eqnarray}
where
\begin{eqnarray}
\nonumber&&{\cal O}^{(5)}=\{\eta^3,\eta\phi^2,\theta^3\}\\
\nonumber&&{\cal
O}^{(6)}=\{\varphi^2\eta,\varphi^2\phi,\chi^2\eta,\chi^2\phi,\varphi\chi\eta,\varphi\chi\phi\}\\
\label{40}&&{\cal
O}^{(7)}=\{\varphi^2\theta\Delta,\chi^2\theta\Delta,\eta^4,\eta^2\Delta^2,\eta^2\phi^2,\eta\phi^3,\phi^4,\phi^2\Delta^2,\Delta^4\}
\end{eqnarray}
Since the quantum numbers of $b^{c}$ and $s^{c}$ are exactly the
same, as is obvious from Table \ref{tab:trans}, there are no
fundamental distinctions between $b^{c}$ and $s^{c}$, we have
defined $b^{c}$ as the one which couples to $Q_3\theta h_d$ in the
superpotential $w_d$. We notice that both the supermultiplets
$\varphi$, $\chi$ and $\eta$, $\phi$, which control the flavor
symmetry breaking in the charged lepton and neutrino sectors
respectively, couple to the quarks. Consequently the $S_4$ flavor
symmetry is completely broken in the quark sector. By recalling the
vacuum configuration in Eq.(\ref{9}) and Eq.(\ref{12}), we can write
down the mass matrices for the up and down quarks
\begin{eqnarray}
\label{41}&&m_{u}=\left(\begin{array}{ccc}
y^{(u)}_{11}\frac{v_{\varphi}v^3_{\phi}}{\Lambda^4}&y^{(u)}_{12}\frac{v_{\varphi}v^3_{\phi}}{\Lambda^4}&y^{(u)}_{13}\frac{v^3_{\varphi}}{\Lambda^3}\\
y^{(u)}_{21}\frac{v^2_{\phi}v_{\Delta}}{\Lambda^3}&y^{(u)}_{22}\frac{v^2_{\varphi}}{\Lambda^2}&y^{(u)}_{23}\frac{v_{\theta}}{\Lambda}\\
y^{(u)}_{31}\frac{v^2_{\phi}v_{\Delta}}{\Lambda^3}&y^{(u)}_{32}\frac{v^2_{\varphi}}{\Lambda^2}&y^{(u)}_{33}
\end{array}\right)v_{u}\\
\nonumber&&\\
\label{42}&&m_{d}=\left(\begin{array}{ccc}
y^{(d)}_{11}\frac{v^2_{\varphi}v_{\phi}}{\Lambda^3}&y^{(d)}_{12}\frac{v^2_{\varphi}v_{\phi}}{\Lambda^3}&y^{(d)}_{13}\frac{v^4_{\phi}}{\Lambda^4}\\
y^{(d)}_{21}\frac{v^2_{\phi}v_{\Delta}}{\Lambda^3}&y^{(d)}_{22}\frac{v^2_{\varphi}}{\Lambda^2}&y^{(d)}_{23}\frac{v^3_{\theta}}{\Lambda^3}\\
y^{(d)}_{31}\frac{v^2_{\phi}v_{\Delta}}{\Lambda^3}&y^{(d)}_{32}\frac{v^2_{\varphi}}{\Lambda^2}&y^{(d)}_{33}\frac{v_{\theta}}{\Lambda}
\end{array}\right)v_d
\end{eqnarray}
where $y^{(u)}_{ij}$ and $y^{(d)}_{ij}$ ($i,j=1,2,3$) are the sum of
all the different terms appearing in the superpotential, all of them
are expect to be of order one. We note that the contribution of
$d^{c}Q_3(\varphi\chi)_{1_2}\Delta h_d$ vanishes with the LO vacuum
alignment, accordingly the (13) element of the down quark mass
matrix $m_d$ arise at order $1/\Lambda^4$. Diagonalizing the above
quark mass matrices in Eq.(\ref{41}) and Eq.(\ref{42}) with the
standard perturbation technique, we obtain the quark masses as
follows
\begin{eqnarray}
\nonumber&&m_{u}\simeq\Big|y^{(u)}_{11}\frac{v_{\varphi}v^3_{\phi}}{\Lambda^4}v_u\Big|\\
\nonumber&&m_c\simeq\Big|y^{(u)}_{22}\frac{v^2_{\varphi}}{\Lambda^2}v_u\Big|\\
\nonumber&&m_{t}\simeq\Big|y^{(u)}_{33}v_u\Big|\\
\nonumber&&m_{d}\simeq\Big|y^{(d)}_{11}\frac{v^2_{\varphi}v_{\phi}}{\Lambda^3}v_d\Big|\\
\nonumber&&m_{s}\simeq\Big|y^{(d)}_{22}\frac{v^2_{\varphi}}{\Lambda^2}v_d\Big|\\
\label{43}&&m_b\simeq\Big|y^{(d)}_{33}\frac{v_{\theta}}{\Lambda}v_d\Big|
\end{eqnarray}
We see that the quark mass hierarchies are correctly produced if the
VEVs $v_{\eta}$, $v_{\phi}$, $v_{\theta}$ and $v_{\Delta}$ are of
order ${\cal O}(\lambda^2_c\Lambda)$ as well. This is consistent
with our naive expectation that all the VEVs should be of the same
order of magnitude. Note that the quark mass hierarchies are
generated through the spontaneous breaking of the flavor symmetry
instead of the FN mechanism. It is obvious that the mass hierarchies
between top and bottom quark mainly come from the symmetry breaking
parameter $v_{\theta}/\Lambda$, and $\tan\beta\equiv\frac{v_u}{v_d}$
should be of order one in our model. Comparing with the tau lepton
mass $m_{\tau}$ predicted in Eq.(\ref{18}), we see that $m_{\tau}$
and $m_b$ are of the same order, this is consistent with the
$b-\tau$ unification predicted in many grand unification models.

For the quark mixing, the CKM matrix elements are estimated as
\begin{eqnarray}
\nonumber&&V_{ud}\simeq V_{cs}\simeq V_{tb}\simeq1\\
\nonumber&&V^{*}_{us}\simeq-V_{cd}\simeq\Big(\frac{y^{(d)}_{21}}{y^{(d)}_{22}}-\frac{y^{(u)}_{21}}{y^{(u)}_{22}}\Big)\frac{v^3_{\phi}}{\Lambda
v^2_{\varphi}}\\
\nonumber&&V^{*}_{ub}\simeq\frac{y^{(u)}_{22}y^{(d)}_{31}-y^{(u)}_{21}y^{(d)}_{32}}{y^{(u)}_{22}y^{(d)}_{33}}\frac{v^3_{\phi}}{\Lambda^2v_{\theta}}\\
\nonumber&&V^{*}_{cb}\simeq-V_{ts}\simeq\frac{y^{(d)}_{32}}{y^{(d)}_{33}}\frac{v^2_{\varphi}}{\Lambda
v_{\theta}}\\
\label{44}&&V_{td}\simeq\frac{y^{(d)}_{21}y^{(d)}_{32}-y^{(d)}_{22}y^{(d)}_{31}}{y^{(d)}_{22}y^{(d)}_{33}}\frac{v^3_{\phi}}{\Lambda^2v_{\theta}}
\end{eqnarray}
We see that the correct orders of the CKM matrix elements are reproduced with
the exception of Cabibbo angle. The $V_{us}$(or $V_{cd}$) is the
combination of two independent contributions of order $\lambda^2_c$,
we need an accidental enhancement of the combination
$\Big(\frac{y^{(d)}_{21}}{y^{(d)}_{22}}-\frac{y^{(u)}_{21}}{y^{(u)}_{22}}\Big)$
of order $1/\lambda_c$ in order to obtain the correct Cabibbo angle.

\section{Phenomenological implications}
In the following we shall study the constraints on the model imposed by the observed
values of $\Delta m^2_{sol}\equiv m^2_2-m^2_1$ and $\Delta
m^2_{atm}\equiv|m^2_3-m^2_1(m^2_2)|$. The important physical consequences
of our model are investigated in details, and the corresponding
predictions are presented. In this section we mainly concentrate on
the neutrino sector. We assume that the right handed neutrino mass
$M$ is much smaller than the cutoff scale $\Lambda$ of the theory,
then the light neutrino masses are dominantly generated via the
See-Saw mechanism.
\subsection{The neutrino mass spectrum}
According to Eq.(\ref{28}), the light neutrino mass spectrum is
controlled by two parameters $a$ and $b$, which are in general both
complex numbers. For convenience, we define
\begin{equation}
\label{45}\frac{b}{a}=R\,e^{i\Phi}
\end{equation}
with $R=|\frac{b}{a}|$. As we will see in the following, all the low
energy observables can be expressed in terms of only three
independent quantities: the ratio $R$, the relative phase $\Phi$
between $a$ and $b$, and the lightest neutrino mass. Experimentally,
only two spectrum observables $\Delta m^2_{sol}$ and $\Delta
m^2_{atm}$ have been measured, therefore the light neutrino mass
spectrum can be normal hierarchy(NH) or inverted hierarchy(IH). The
ratio between $\Delta m^2_{sol}$ and $\Delta m^2_{atm}$ is given by
\begin{equation}
\label{add1}\frac{\Delta m^2_{sol}}{\Delta
m^2_{atm}}=\frac{15-81R^4-18R^2-36R^2\cos^2\Phi+12R(1+9R^2)\cos\Phi}{24R(1+9R^2)|\cos\Phi|}
\end{equation}
where we have taken $\Delta m^2_{atm}=|m_3^2-m_1^2|$ for both the NH
and IH neutrino spectrums for convenience. Moreover, we have the
following relationships for the neutrino masses
\begin{eqnarray}
\nonumber&&\frac{16m^2_1}{m^2_2}=1+81R^4+18R^2+36R^2\cos^2\Phi-12R(1+9R^2)\cos\Phi\\
\label{46}&&\frac{16m^2_3}{m^2_2}=1+81R^4+18R^2+36R^2\cos^2\Phi+12R(1+9R^2)\cos\Phi
\end{eqnarray}
Then the parameters $R$ and $\cos\Phi$ can be expressed in terms of
light neutrino mass as follows
\begin{equation}
\label{47}\left\{\begin{array}{l}R=\frac{1}{3}\sqrt{\frac{2(m_3+m_1)}{m_2}-1}\\
\\
\cos\Phi=\frac{m_3-m_1}{m_2}\frac{1}{\sqrt{\frac{2(m_3+m_1)}{m_2}-1}}
\end{array}\right.
\end{equation}
or
\begin{equation}
\label{48}\left\{\begin{array}{l}
R=\frac{1}{3}\sqrt{\frac{2|m_3-m_1|}{m_2}-1}\\
\\
\cos\Phi=\frac{m^2_3-m^2_1}{m_2|m_3-m_1|}\frac{1}{\sqrt{\frac{2|m_3-m_1|}{m_2}-1}}
\end{array}\right.
\end{equation}

\begin{figure}
\begin{center}
\begin{tabular}{cc}
\includegraphics[scale=.745]{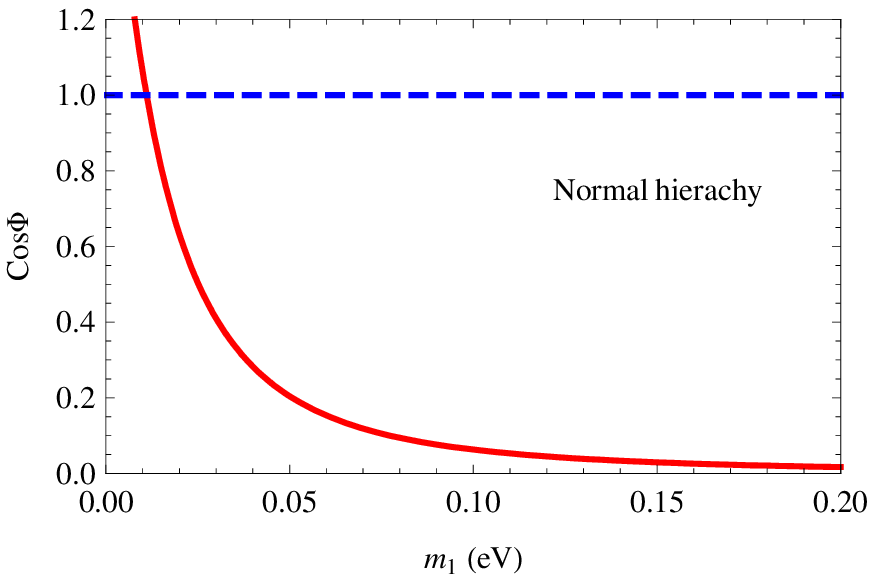}&\includegraphics[scale=.745]{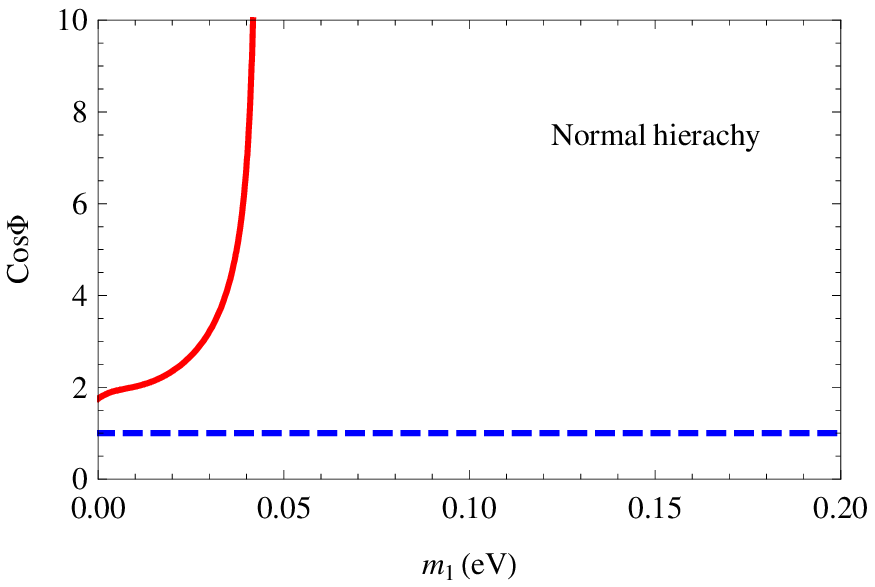}\\
(a)&(b)
\end{tabular}
\caption{\label{fig:cosphi_No} The variation of $\cos\Phi$ with
respect to the lightest neutrino mass $m_1$ for the normal hierarchy
spectrum. In Fig. \ref{fig:cosphi_No}a, $\cos\Phi$ is the taken to
be the expression in Eq.(\ref{47}), and Fig. \ref{fig:cosphi_No}b
corresponds to the value of $\cos\Phi$ in Eq.(\ref{48}).}
\end{center}
\end{figure}
\begin{figure}
\begin{center}
\begin{tabular}{cc}
\includegraphics[scale=.745]{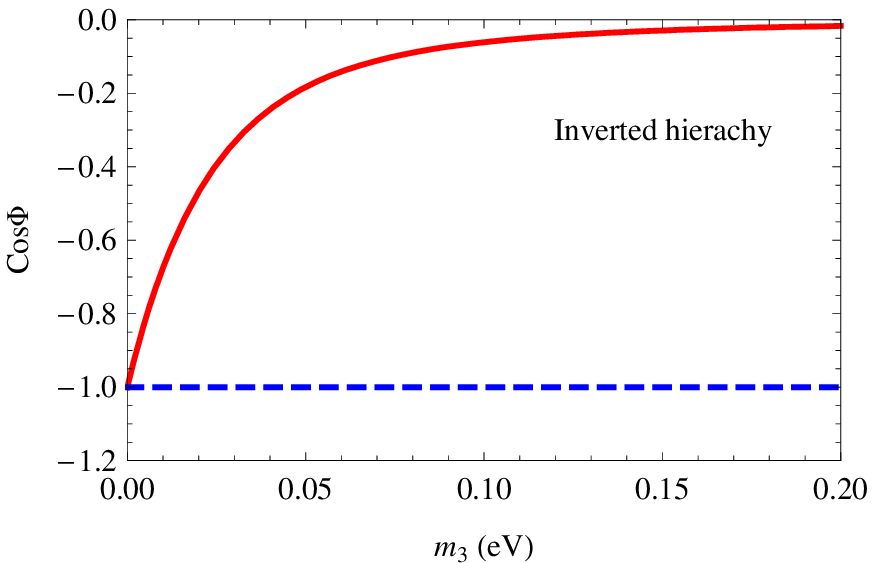}&\includegraphics[scale=.745]{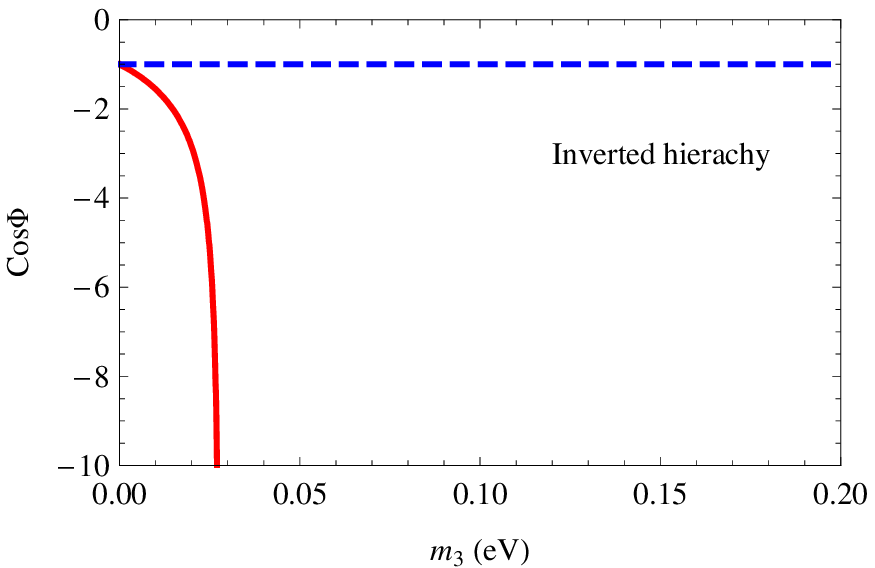}\\
(a)&(b)
\end{tabular}
\caption{\label{fig:cosphi_Io} $\cos\Phi$ as a function of the
lightest neutrino mass $m_3$ for the inverted hierarchy spectrum.
Fig. \ref{fig:cosphi_Io}a and  Fig. \ref{fig:cosphi_Io}b are for the
$\cos\Phi$ values in Eq.(\ref{47}) and Eq.(\ref{48}) respectively.}
\end{center}
\end{figure}

These results hold for both the normal hierarchy and inverted
hierarchy spectrum. In the case of normal hierarchy, $m_2$ and $m_3$
can be expressed as functions of the lightest neutrino mass:
$m_2=\sqrt{m^2_1+\Delta m^2_{sol}}$ and $m_3=\sqrt{m^2_1+\Delta
m^2_{atm}}$. For the inverted hierarchy, $m_3$ is the lightest
neutrino mass, the remaining two masses are $m_1=\sqrt{m^2_3+\Delta
m^2_{atm}}$ and $m_2=\sqrt{m^2_3+\Delta m^2_{sol}+\Delta
m^2_{atm}}$. As a result, taking into account the experimental
information on $\Delta m^2_{sol}$ and $\Delta m^2_{atm}$, there is
only one real parameter undetermined, and it is chose to be the
lightest neutrino mass $m_l$($m_1$ or $m_3$) in the present work.
Hence our model is quite predictive. We display $\cos\Phi$ as a
function of the lightest neutrino mass in Fig.\ref{fig:cosphi_No}
and Fig.\ref{fig:cosphi_Io} for the normal hierarchy and inverted
hierarchy respectively, where the best fit values of $\Delta
m^2_{sol}=7.65\times10^{-5}\;{\rm eV^2}$ and $\Delta
m^2_{atm}=2.40\times10^{-3}\;{\rm eV^2}$ have been used. For the
solution of $R$ and $\cos\Phi$ shown in Eq.(\ref{48}), we can
clearly see that the corresponding value of $|\cos\Phi|$ would be
larger than 1 in the case of both normal hierarchy and inverted
hierarchy spectrum. Furthermore, we have verified that $|\cos\Phi|$
is always larger than 1 for the $3\sigma$ range of $\Delta
m^2_{sol}$ and $\Delta m^2_{atm}$, therefore the solution in
Eq.(\ref{48}) can be disregarded thereafter. From the condition
$|\cos\Phi|\leq1$, we obtain the following constraints on the
lightest neutrino mass,
\begin{eqnarray}
\nonumber&&m_1\geq0.011\;{\rm eV},~~~~{\rm Normal~~~hierarchy}\\
\label{49}&&m_3>0.0\;{\rm eV},~~~~~~~{\rm Inverted~~~hierarchy}
\end{eqnarray}

For the NH spectrum, we have a lower bound on $m_1$, which is
satisfied for $\Phi=0$. The corresponding values of $\cos\Phi$ are positive, then
$\Phi$ is in the range of $0\sim\pi/2$ or $3\pi/2\sim2\pi$. In the
case of IH, $\cos\Phi$ is negative so that $\Phi$ varies between
$\pi/2$ to $3\pi/2$. From Fig. \ref{fig:cosphi_Io} we can see that
$\cos\Phi$ is very close to -1 for $m_3$ tending to zero, the
lightest neutrino mass $m_3$ is less constrained.

\subsection{Neutrinoless double beta decay}
\begin{figure}[hptb]
\begin{center}
\includegraphics[scale=.745]{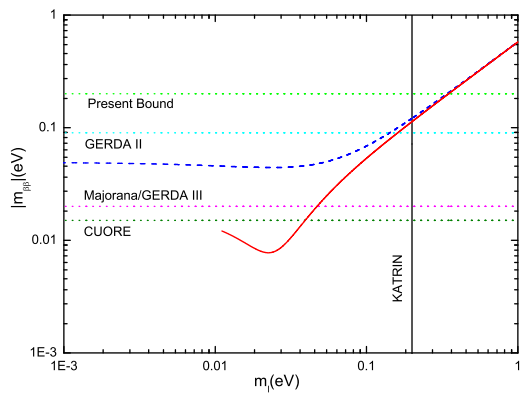}
\caption{\label{fig:0nubeta} $m_{\beta\beta}$ as a function of the lightest neutrino mass $m_l$, the solid and dashed lines represent the NH and IH cases respectively.}
\end{center}
\end{figure}

Neutrinoless double beta decay (0$\nu2\beta$) is a sensitive probe
to the scale of the neutrino masses, it is a very slow
lepton-number-violating nuclear transition that occurs if neutrinos
have mass and are Majorana particles. The rate of $0\nu2\beta$ decay
is determined by the nuclear matrix elements and the effective
0$\nu2\beta$-decay mass $|m_{\beta\beta}|$, which is defined as
$|m_{\beta\beta}|=|\sum_{k}(U_{PMNS})^2_{ek}m_k|$. In the present
model it is given by
\begin{eqnarray}
\nonumber&&|m_{\beta\beta}|=|2a^2+6b^2-4ab|\frac{v^2_{u}}{|M|}\\
\label{50}&&~~~=\frac{m_2}{2}\Big[1+9R^4+4R^2+6R^2\cos(2\Phi)-12R^3\cos\Phi-4R\cos\Phi\Big]^{1/2}
\end{eqnarray}
By using Eq.(\ref{47}), we can express $|m_{\beta\beta}|$ in terms
of the lightest neutrino mass $m_l$, the corresponding results are
shown in Fig.\ref{fig:0nubeta}. The vertical line represents the
future sensitivity of the KATRIN experiment \cite{katrin}, the
horizontal ones denote the present bound from the Heidelberg-Moscow
experiment \cite{HM} and the future sensitivity of some $0\nu2\beta$
decay experiments, which are 15 meV, 20 meV and 90 meV, respectively
of CUORE \cite{cuore}, Majorana \cite{majorana}/GERDA III
\cite{gerda} and GERDA II experiments. From Fig. \ref{fig:0nubeta}
we conclude that for the allowed values of $m_l$, the predictions
for $m_{\beta\beta}$ approach the future experimental sensitivity.
For the NH spectrum, the effective mass $m_{\beta\beta}$ can reach a
very low value about 7.8 meV. Whereas the lower bound for
$m_{\beta\beta}$ is approximately 44.3 meV in the case of IH. A
combined measurement of the effective mass $m_{\beta\beta}$ and the
lightest neutrino mass can determine whether the neutrino spectrum
is NH or IH in our model.

\subsection{Beta decay}
\begin{figure}
\begin{center}
\includegraphics[scale=.745]{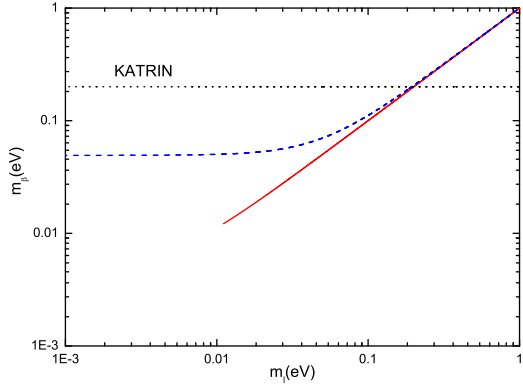}
\caption{\label{fig:beta} Variation of $m_{\beta}$ with respect to the lightest neutrino mass $m_l$, the solid and dashed lines represent the NH and IH spectrum respectively.}
\end{center}
\end{figure}

One can directly search for the kinetic effect of nonzero neutrino
masses in beta decay by modification of the Kurie plot. This search
is sensitive to neutrino masses regardless of whether the neutrinos
are Dirac or Majorana particles. For small neutrino masses, this
effect will occur near to the end point of the electron energy
spectrum and will be sensitive to the quantity
$m_{\beta}=\big[\sum_k|(U_{PMNS})_{ek}|^2m^2_k\big]^{1/2}$. For the
present model, we have
\begin{equation}
\label{51}m_{\beta}=\frac{1}{\sqrt{3}}(2m^2_1+m^2_2)^{1/2}
\end{equation}
This result holds for both the NH and IH spectrums. In Fig.\ref{fig:beta} we plot $m_{\beta}$ versus the lightest neutrino mass $m_{l}$, the horizontal line represents the future sensitivity of 0.2 eV from the KATRIN experiment.

\subsection{Sum of the neutrino masses}
\begin{figure}[hptb]
\begin{center}
\includegraphics[scale=.745]{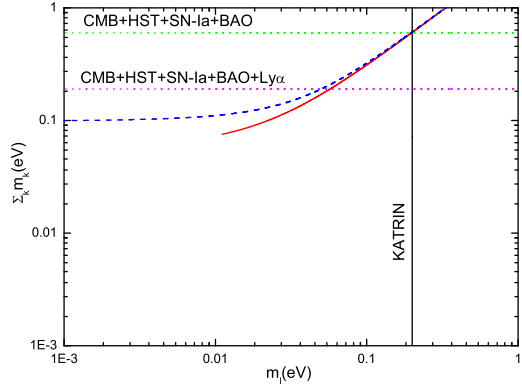}
\caption{\label{fig:sum} The sum of the neutrino masses $\sum_km_k$ versus the lightest neutrino mass $m_l$, the solid and dashed lines
represent the NH and IH spectrum respectively.}
\end{center}
\end{figure}

The sum of the neutrino masses $\sum_km_k$ is constrained by the
cosmological observation. In Fig. \ref{fig:sum} we display the sum
of the neutrino masses as a function of the lightest neutrino mass
$m_l$. The vertical line denotes the future sensitivity of KATRIN
experiment, and the horizontal lines are the cosmological bounds
\cite{Fogli:2008cx}. There are typically five representative
combinations of the cosmological data, which lead to increasingly
stronger upper bounds on the sum of the neutrino masses. We show the
two strongest ones in Fig. \ref{fig:sum}. The first one at $0.60$ eV
corresponds to the combination of the Cosmic Microwave Background
(CMB) anisotropy data (from WMAP~5y \cite{WMAP2}, Arcminute
Cosmology Bolometer Array Receiver (ACBAR) \cite{acbar07}, Very
Small Array (VSA) \cite {vsa}, Cosmic Background Imager (CBI)
\cite{cbi} and BOOMERANG \cite{boom03} experiments) plus the
large-scale structure (LSS) information on galaxy clustering (from
the Luminous Red Galaxies Sloan Digital Sky Survey (SDSS)
\cite{Tegmark}) plus the Hubble Space Telescope (HST) plus the
luminosity distance SN-Ia data of \cite{astier} and finally plus the
BAO data from \cite{bao}. The second one at $0.19$ eV corresponds to
all the previous data combined to the small scale primordial
spectrum from Lyman-alpha (Ly$\alpha$) forest clouds \cite{Ly1}. We
see that the current cosmological information on the sum of the
neutrino masses can hardly distinguish the NH spectrum from the IH
spectrum.

\subsection{The Majorana CP violating phases}
\begin{figure}[hptb]
\begin{center}
\begin{tabular}{cc}
\includegraphics[scale=.745]{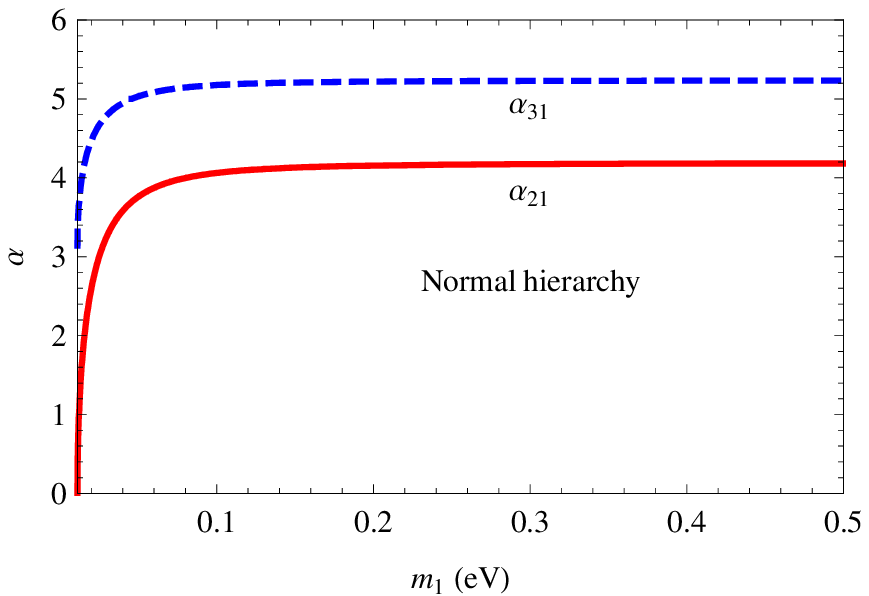}&\includegraphics[scale=.745]{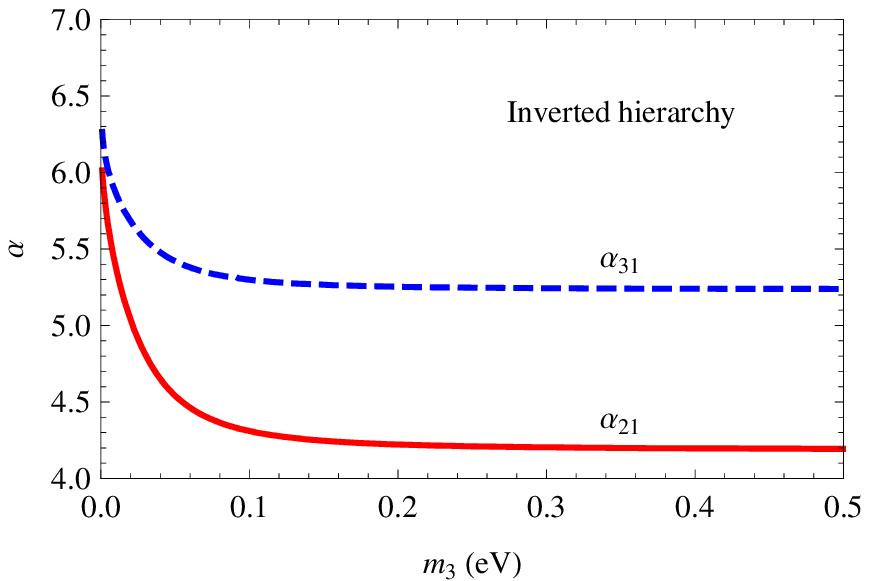}\\
(a)&(b)
\end{tabular}
\caption{\label{fig:majorana_phase} The dependence of the Majorana
CP violating phases $\alpha_{21}$ and $\alpha_{31}$ on the lightest
neutrino mass $m_l$. Solid and dashed lines refer to $\alpha_{21}$
and $\alpha_{31}$ respectively. Fig.\ref{fig:majorana_phase}a
corresponds to the NH mass spectrum, and
Fig.\ref{fig:majorana_phase}b is for the IH case, where $\sin\Phi$
is taken to be positive.}
\end{center}
\end{figure}

In the standard parametrization \cite{pdg}, the lepton PMNS mixing matrix is defined by
\begin{equation}
\label{52}U_{PMNS}=\left(\begin{array}{ccc}c_{12}c_{13}&s_{12}c_{13}&s_{13}e^{-i\delta}\\
-s_{12}c_{23}-c_{12}s_{23}s_{13}e^{i\delta}&c_{12}c_{23}-s_{12}s_{23}s_{13}e^{i\delta}&s_{23}c_{13}\\
s_{12}s_{23}-c_{12}c_{23}s_{13}e^{i\delta}&-c_{12}s_{23}-s_{12}c_{23}s_{13}e^{i\delta}&c_{23}c_{13}\end{array}\right)\,{\rm diag}(1,e^{i\alpha_{21}/2},e^{i\alpha_{31}/2})
\end{equation}
where $c_{ij}=\cos\theta_{ij}$, $s_{ij}=\sin\theta_{ij}$ with $\theta_{ij}\in[0,\pi/2]$, $\delta$ is the Dirac CP violating phase, $\alpha_{21}$ and $\alpha_{31}$ are the two
Majorana CP violating phases, all the three CP violating phases $\delta$, $\alpha_{21}$ and $\alpha_{31}$ are allowed to vary in the range of $0\sim2\pi$. Recalling that the leptonic
mixing matrix is given by Eq.(\ref{26}) at LO, in the standard parametrization it is,
\begin{equation}
\label{53}U_{PMNS}=e^{-i\alpha_1/2}\;{\rm diag}(1,1,-1)U_{TB}\;{\rm diag}(1,e^{i(\alpha_1-\alpha_2)/2},e^{i(\alpha_1-\alpha_3)/2})
\end{equation}
where the overall phase $e^{-i\alpha_1/2}$ can be absorbed into the charged lepton fields. Comparing Eq.(\ref{53}) with Eq.(\ref{52}), we can identify the two CP
violating phases as
\begin{equation}
\label{54}\alpha_{21}=\alpha_1-\alpha_2,~~~~~~~\alpha_{31}=\alpha_{1}-\alpha_3
\end{equation}
Similar to other low energy observables, $\alpha_{21}$ and $\alpha_{31}$ can be written as functions of the parameters $R$ and $\Phi$,
\begin{eqnarray}
\label{55}&&\left\{\begin{array}{l}\sin\alpha_{21}=\frac{9R^2\sin(2\Phi)-6R\sin\Phi}{1+9R^2-6R\cos\Phi},\\
\\
\cos\alpha_{21}=\frac{1+9R^2\cos(2\Phi)-6R\cos\Phi}{1+9R^2-6R\cos\Phi}\end{array}\right.\\
\nonumber&&\\
\label{56}&&\left\{\begin{array}{l}\sin\alpha_{31}=\frac{12R(1-9R^2)\sin\Phi}{1+81R^4+18R^2-36R^2\cos^2\Phi}\\
\\
\cos\alpha_{31}=-\frac{1+81R^4-36R^2+18R^2\cos(2\Phi)}{1+81R^4+18R^2-36R^2\cos^2\Phi}
\end{array}\right.
\end{eqnarray}
Note that the relations between  $R$, $\Phi$ and the light neutrino
masses are displayed in Eq.(\ref{47}). In contrast to other low
energy observables such as the light neutrino masses,
$m_{\beta\beta}$ and $m_{\beta}$ etc., the Majorana phases
$\alpha_{21}$ and $\alpha_{31}$ depend on both $\cos\Phi$ and
$\sin\Phi$, and not only on $\cos\Phi$. In
Fig.\ref{fig:majorana_phase} we show the behavior of the Majorana
phases $\alpha_{21}$ and $\alpha_{31}$ with respect to the lightest
neutrino mass $m_{l}$, where we choose $\sin\Phi>0$ for
illustration.

It is well-known that the See-Saw mechanism provides an elegant
explanation for the smallness of the neutrino mass, meanwhile the
baryon asymmetry may be produced through the out of equilibrium CP
violating decays of the right handed neutrino $\nu^{c}$. As is shown
in Eq.(\ref{23}), at LO the heavy neutrino masses are exactly
degenerate in our model, then leptogenesis can be naturally
implemented via the so-called resonant leptogenesis mechanism
\cite{Pilaftsis:2005rv}. Since more other subtle issues are involved
in the resonant leptogenesis, the analysis of whether the observed
baryon asymmetry can be naturally generated in our model is beyond
the range of the present paper, which will be discussed in future
work \cite{ding}.

The phenomenological consequences for the LO order predictions of
our model have been analyzed . As we shall show in the next section
that our model gets corrections when higher dimensional operators
are included in the Lagrangian. These corrections modify the leading
order predictions by terms of relative order $\lambda^2_c$, hence
the results presented so far are still correct approximately.
However, we would like to note that due to the next to the leading
order contributions, a cancallation could be present for the
effective $0\nu2\beta$-decay mass, consequently $m_{\beta\beta}$
could reach zero in the NH case.

\section{Next to the leading order corrections}
The results of the previous section hold to first approximation. At
the next to leading order (NLO), the superpotentials $w_{v}$,
$w_{\ell}$, $w_{\nu}$, $w_{u}$ and $w_{d}$ are corrected by higher
dimensional operators compatible with the symmetry of the model,
whose contributions are suppressed by at least one additional power
of $\Lambda$. The residual Klein four and $Z_3$ symmetry in the
neutrino and the charged lepton sectors at LO would be broken
completely by the NLO contributions. The NLO terms in the driving
superpotential leads to small deviation from the LO vacuum
alignment. The masses and mixing matrices are corrected by both the
shift of the vacuum configuration and the NLO operators in the
Yukawa superpotentials $w_{\ell}$, $w_{\nu}$, $w_{u}$ and $w_{d}$.
In the following, the NLO corrections to the vacuum alignment and
the mass matrices will be discussed one by one, and the resulting
physical effects are studied.

\subsection{Corrections to the vacuum alignment}

The NLO operators of the driving superpotential $w_v$ and the
corresponding corrections to the LO vacuum alignment in Eq.(\ref{9})
and Eq.(\ref{12}) are discussed in the Appendix B in details. The
inclusion of the higher dimensional operators results in the shift
of the VEVs of the flavon fields, the vacuum configuration is
modified into
\begin{eqnarray}
\nonumber&&\langle\varphi\rangle=(\delta v_{\varphi_1},v_{\varphi}+\delta v_{\varphi_2},\delta v_{\varphi_3}),~~~~\langle\chi\rangle=(\delta v_{\chi_1},v_{\chi},\delta v_{\chi_3}),\\
\nonumber&&\langle\eta\rangle=(v_{\eta},v_{\eta}+\delta v_{\eta_2}),~~~~~~\langle\phi\rangle=(v_{\phi}+\delta v_{\phi},v_{\phi}+\delta v_{\phi},v_{\phi}+\delta v_{\phi}),\\
\label{57}&&\langle\theta\rangle=v_{\theta}+\delta v_{\theta} 
\end{eqnarray}
where $v_{\chi}$ and $v_{\eta}$ are still undetermined, and the VEV
$v_{\Delta}$ is not corrected by the NLO terms. All the corrections
are suppressed by $1/\Lambda$, and the shift of $\langle\phi\rangle$
turns out to be proportional to its LO VEV. Since all the VEVs are
required to be of order ${\cal O}(\lambda^2_c\Lambda)$, we expect
these corrections would modify the LO VEVs by terms of relative
order $\lambda^2_c$.

\subsection{Corrections to the mass matrices}

The corrections to the fermion mass matrices originate from two
sources: the first is the higher dimensional operators in the Yukawa
superpotentials $w_{\ell}$, $w_{\nu}$, $w_{u}$ and $w_{d}$, and the
second is the deviation from the LO vacuum alignment, which is
induced by the NLO terms in the driving potential. As a result, at
NLO the mass matrices are the sum of the contributions of higher
dimension operators evaluated with the insertion of the the LO VEV,
and those from the LO superpotentials evaluated with the NLO VEVs.

For the charged leptons, the superpotential $w_{\ell}$ is corrected
by the following sisteen NLO operators
\begin{eqnarray}
\nonumber&&e^{c}(\ell\varphi^3)_{1_2}\theta
h_d,~~~~e^{c}(\ell\varphi^2\chi)_{1_2}\theta
h_d,~~~~e^{c}(\ell\varphi\chi^2)_{1_2}\theta
h_d,~~~e^{c}(\ell\chi^3)_{1_2}\theta h_d,\\
\nonumber&&e^{c}(\ell\varphi\eta^2)_{1_2}\Delta h_d,~~e^{c}(\ell\chi\eta^2)_{1_2}\Delta h_d,~~e^{c}(\ell\varphi\eta\phi)_{1_2}\Delta h_d,~~e^{c}(\ell\chi\eta\phi)_{1_2}\Delta h_d,\\
\nonumber&&e^{c}(\ell\varphi\phi^2)_{1_2}\Delta h_d,~~e^{c}(\ell\chi\phi^2)_{1_2}\Delta h_d,~~\mu^{c}(\ell\varphi^2)_{1_1}\theta h_d,~~~~~\mu^{c}(\ell\chi^2)_{1_1}\theta h_d,\\
\label{58}&&\mu^{c}(\ell\varphi\chi)_{1_1}\theta h_d,~~~~\mu^{c}(\ell\phi^2)_{1_1}\Delta h_d,~~~\mu^{c}(\ell\eta\phi)_{1_1}\Delta h_d,~~~~\tau^{c}(\ell\chi)_{1_2}\theta h_d
\end{eqnarray}
Taking into account the contributions of the modified vacuum
alignment at NLO, each diagonal entry of the charged lepton mass
matrix receives a small correction factor, while the off-diagonal
entries become non-zero and of the order of the diagonal term in
each row multiplied by $\varepsilon$, which parameterizes the ratio
$VEV/\Lambda$ with order ${\cal O}(\lambda^2_c)$. Then we have
\begin{equation}
\label{59}m_{\ell}=\left(\begin{array}{ccc}
m^{\ell}_{11}\varepsilon^2&m^{\ell}_{12}\varepsilon^3&m^{\ell}_{13}\varepsilon^3\\
m^{\ell}_{21}\varepsilon^2&m^{\ell}_{22}\varepsilon&m^{\ell}_{23}\varepsilon^2\\
m^{\ell}_{31}\varepsilon&m^{\ell}_{32}\varepsilon&m^{\ell}_{33}
\end{array}\right)\varepsilon v_d
\end{equation}
where the coefficients $m^{\ell}_{ij}$($i,j=$1,2,3) are order one
unspecified constants. The hermitian matrix
$m^{\dagger}_{\ell}m_{\ell}$ is diagonalized by the unitary matrix
$U_{\ell}$, which exactly corresponds to the transformation of the
charged leptons used to diagonalize $m_{\ell}$,
\begin{equation}
\label{60}U^{\dagger}_{\ell}m^{\dagger}_{\ell}m_{\ell}U_{\ell}\simeq{\rm diag}(|m^{\ell}_{11}\varepsilon^3|^2,|m^{\ell}_{22}\varepsilon^2|^2,|m^{\ell}_{33}\varepsilon|^2)v^2_d
\end{equation}
The charged lepton masses are modified by terms of relative order
$\varepsilon$ with respect to LO results, consequently the NLO
corrections don't spoil the charged lepton mass hierarchies
predicted at LO. The unitary matrix $U_{\ell}$ is approximately
given by
\begin{equation}
\label{61}U_{\ell}\simeq\left(\begin{array}{ccc}
1&(\frac{m^{\ell}_{21}}{m^{\ell}_{22}}\varepsilon)^{*}&(\frac{m^{\ell}_{31}}{m^{\ell}_{33}}\varepsilon)^{*}\\
-\frac{m^{\ell}_{21}}{m^{\ell}_{22}}\varepsilon&1&(\frac{m^{\ell}_{32}}{m^{\ell}_{33}}\varepsilon)^{*}\\
-\frac{m^{\ell}_{31}}{m^{\ell}_{33}}\varepsilon&-\frac{m^{\ell}_{32}}{m^{\ell}_{33}}\varepsilon&1
\end{array}\right)
\end{equation}
Then we turn to the neutrino sector. The NLO correction to the
Majorana masses of the right handed neutrino arises at order
$1/\Lambda$, the corresponding higher dimensional operator is
$(\nu^{c}\nu^{c})_{1_1}\theta^2$, whose contribution can be
completely absorbed into the redefinition of the mass parameter $M$.
The NLO corrections to the neutrino Dirac couplings are
\begin{eqnarray}
\label{62}\frac{y_{\nu1}}{\Lambda}(\nu^{c}\ell\delta\eta)_{1_1}h_u+\frac{y_{\nu2}}{\Lambda}(\nu^{c}\ell\delta\phi)_{1_1}h_u+\frac{x_{\nu1}}{\Lambda^2}(\nu^{c}\ell\eta)_{1_2}\theta h_u+\frac{x_{\nu2}}{\Lambda^2}(\nu^{c}\ell\phi)_{1_2}\theta h_u
\end{eqnarray}
where $\delta\eta$ and $\delta\phi$ represent the shifted VEVs of
the flavons $\eta$ and $\phi$ respectively. Through redefining the
LO parameters $a\rightarrow
a-x_{\nu1}\frac{v_{\eta}v_{\theta}}{\Lambda^2}$ and $b\rightarrow
b+y_{\nu2}\frac{\delta v_{\phi}}{\Lambda}$, the NLO corrections to
$m^D_{\nu}$ are
\begin{equation}
\label{63}\delta m^D_{\nu}=\left(\begin{array}{ccc}
0&\delta_2&\delta_1-\delta_2\\
-\delta_2&\delta_1&\delta_2\\
\delta_1+\delta_2&-\delta_2&0
\end{array}\right)v_u
\end{equation}
where $\delta_1=y_{\nu1}\frac{\delta
v_{\eta_2}}{\Lambda}+2x_{\nu1}\frac{v_{\eta}v_{\theta}}{\Lambda^2}$
and $\delta_2=x_{\nu2}\frac{v_{\phi}v_{\theta}}{\Lambda^2}$. We
notice that both $\delta_1$ and $\delta_2$ are of order $\varepsilon
a$ (or $\varepsilon b$). Therefore, the NLO corrections to the light
neutrino mass matrix are given by
\begin{eqnarray}
\nonumber&&\delta m_{\nu}=-(m^{D}_{\nu})^{T}M^{-1}_{N}\delta m^{D}_{\nu}-(\delta m^{D}_{\nu})^{T}M^{-1}_{N}m^{D}_{\nu}\\
\label{64}&&~~~=\frac{v^2_u}{M}\left(\begin{array}{ccc}-2(a-b)\delta_1&-(2a+b)\delta_1-6b\delta_2&-b(\delta_1-6\delta_2)\\
-(2a+b)\delta_1-6b\delta_2&2b(\delta_1+3\delta_2)&-(2a+b)\delta_1\\
-b(\delta_1-6\delta_2)&-(2a+b)\delta_1&-2(a-b)\delta_1-6b\delta_2
\end{array}\right)
\end{eqnarray}
Diagonalizing the modified light neutrino mass matrix, we obtain the neutrino masses to LO in $\delta_{1,2}$ as follows,
\begin{eqnarray}
\nonumber&&m_1=\Big|(a-3b)^2+(a-3b)\delta_1\Big|\frac{v^2_{u}}{|M|}\\
\nonumber&&m_{2}=4\Big|a^2+a\delta_1\Big|\frac{v^2_u}{|M|}\\
\label{65}&&m_{3}=\Big|(a+3b)^2+(a+3b)\delta_1\Big|\frac{v^2_u}{|M|}
\end{eqnarray}
The PMNS matrix becomes $U_{PMNS}=U^{\dagger}_{\ell}U_{\nu}$, where
$U_{\ell}$ associated with the diagonalization of the charged lepton
mass matrix is given by Eq.(\ref{61}), and the unitary matrix
$U_{\nu}$ diagonalizes the neutrino mass matrix $m_{\nu}+\delta
m_{\nu}$ including the NLO contributions. The parameters of the
lepton mixing matrix are modified as
\begin{eqnarray}
\nonumber&&|U_{e3}|=\frac{1}{\sqrt{2}}\Big|\frac{1}{6(|a|^2+9|b|^2)(ab^{*}+a^{*}b)}[(a+3b)^2(a^{*}\delta^{*}_1
+6b^{*}\delta^{*}_2)-(a^{*}-3b^{*})^2(a\delta_1+6b\delta_2)]\\
\nonumber&&~~~~+\Big(\frac{m^{\ell}_{21}}{m^{\ell}_{22}}\varepsilon\Big)^{*}-\Big(\frac{m^{\ell}_{31}}{m^{\ell}_{33}}\varepsilon\Big)^{*}\Big|\\
\nonumber&&\sin^2\theta_{12}=\frac{1}{3}\Big[1-\frac{m^{\ell}_{21}}{m^{\ell}_{22}}\varepsilon-\frac{m^{\ell}_{31}}{m^{\ell}_{33}}\varepsilon-\Big(\frac{m^{\ell}_{21}}{m^{\ell}_{22}}\varepsilon\Big)^{*}-\Big(\frac{m^{\ell}_{31}}{m^{\ell}_{33}}\varepsilon\Big)^{*}\Big]\\
\nonumber&&\sin^2\theta_{23}=\frac{1}{2}+\frac{1}{2(|a|^{2}+9|b|^2)(ab^{*}+a^{*}b)}\Big[ab(a^{*}\delta^{*}_1+6b^{*}\delta^{*}_2)+a^{*}b^{*}(a\delta_1+6b\delta_2)\Big]\\
\label{66}&&~~~~+\frac{1}{2}\Big[\frac{m^{\ell}_{32}}{m^{\ell}_{33}}\varepsilon+\Big(\frac{m^{\ell}_{32}}{m^{\ell}_{33}}\varepsilon\Big)^{*}\Big]
\end{eqnarray}

We see that the neutrino masses and mixing angles receive
corrections of order $\lambda^2_c$ with respect to LO results. The
value of $\sin^2\theta_{12}$ is still within the $3\sigma$ range of
global data fit, and the corrections to both $\theta_{23}$ and
$\theta_{13}$ are within the current data uncertainties as well. In
particular, a non-vanishing $\theta_{13}$ of order $\lambda^2_c$ is
close to the reach of the next generation neutrino oscillation
experiments and will provide a valuable test of model.

The NLO corrections to the mass matrices in quark sector have been
analyzed following the same method as that for the lepton sector.
Since every entry of the mass matrices $m_u$ and $m_d$ in
Eq.(\ref{41}) and Eq.(\ref{42}) is nonvanishing, the NLO
contributions leads to small corrections of relative order
$\lambda^2_c$ in each entry. Consequently the quark masses and
mixing angles are corrected by terms of relative order $\lambda^2_c$
with respect to LO results, the successful LO predictions are not
spoiled.

\section{Conclusion}

We have constructed a SUSY model for fermion masses and flavor
mixings based on the flavor symmetry $S_4\times Z_3\times Z_4$, the
neutrino masses are assumed to be generated through the See-Saw
mechanism. At LO the $S_4$ symmetry is broken down to Klein four and
$Z_3$ symmetry in the neutrino and charged lepton sector
respectively, this breaking chain exactly leads to the TB mixing. It
is remarkable that the mass hierarchies among the charged leptons
are controlled by the spontaneous breaking of the flavor symmetry.
We further extend the flavor symmetry to the quark sector, where the
$S_4$ symmetry is completely broken. The correct orders of quark
masses and CKM matrix elements are generated with the exception of
the mixing angle between the first two generations, which requires a
samll accidental enhancement.

We have carefully analyzed the NLO contributions due to higher
dimensional operators which modify both the Yukawa couplings and the
LO vacuum alignment, and we have verified that all the fermino
masses and mixing angles are corrected by terms of relative order
$\lambda^2_c$ with respect to the LO results. As a result, the
successful LO predictions are not spoiled. Particularly we expect
the mixing angle $\theta_{13}$ would be of order $\lambda^2_c$, it
is within the sensitivity of the experiments which are now in
preparation and will take data in the near future
\cite{Ardellier:2006mn,Wang:2006ca}. Precise measurement of
$\theta_{13}$ is an important test to our model.

The phenomenological consequences of our model are analyzed in
detail. The low energy observables including the neutrino mass
squared difference, neutrinoless double decay, beta decay, the sum
of the neutrino masses and the Majorana CP violating phases are
considered. All the low energy observables can be expressed in terms
of three independent parameters: the ratio $R=|b/a|$, the relative
phase $\Phi$ between $a$ and $b$ and the lightest neutrino mass
$m_l$. Once the parameters are fixed to match $\Delta m^2_{sol}$ and
$\Delta m^2_{atm}$, there is only one parameter left, which is chose
to be $m_l$ in the present work. Both the normal and inverted
hierarchy neutrino spectrum are allowed in our model. For normal
hierarchy there is a low bound on $m_1$ of approximately 0.011 eV.
In the case of inverted hierarchy, $m_3$ is less constrained and we
only obtain the trivial constraint that $m_3$ should be positive.
The lower bounds of the effective mass $m_{\beta\beta}$ are
approximately 7.8 meV and 44.3 meV for the NH and IH spectrum
respectively. A combined measurement of $m_{\beta\beta}$ and the
lightest neutrino mass can distinguish the NH from the IH spectrum.
The Majorana CP violating phases depend both on $\cos\Phi$ and
$\sin\Phi$, whereas only $\cos\Phi$ is involved in other low energy
observables. It is remarkable that the right handed neutrino masses
are exactly degenerate at LO, the baryon asymmetry may be generated
via the rosonant leptogenesis.

\section*{Acknowledgements}

We are grateful to Prof. Mu-Lin Yan for stimulating discussions.
This work is supported by the Chinese Academy KJCX2-YW-N29 and the
973 project with Grant No. 2009CB825200.

\vfill
\newpage

\section*{Appendix A: Representation matrices of the $S_4$ group}

In this appendix, we explicitly show the representation matrices
of the $S_4$ group for the five irreducible
representations. The matrices for the generators $S$ and $T$ depend
on the representations as follows
\begin{eqnarray*}
\begin{array}{lcc}
1_1,&S=1,&T=1\\
1_2,&S=-1,&T=1\\
2,&~~~S=\left(\begin{array}{cc} 0&1\\
1&0
\end{array}
\right),&~~~T=\left(\begin{array}{cc}
\omega&0\\
0&\omega^2
\end{array}
\right)
\\
3_1,& ~~~~~~~~~S=\frac{1}{3}\left(\begin{array}{ccc} -1&2\omega&2\omega^2\\
2\omega&2\omega^2&-1\\
2\omega^2&-1&2\omega
\end{array}
\right),&~~~~~~~~~ T=\left(\begin{array}{ccc}
1&0&0\\
0&\omega^2&0\\
0&0&\omega
\end{array}
\right)\\
3_2,&~~~~~~~~S=\frac{-1}{3}\left(\begin{array}{ccc} -1&2\omega&2\omega^2\\
2\omega&2\omega^2&-1\\
2\omega^2&-1&2\omega
\end{array}
\right), &~~~~~~~~~ T=\left(\begin{array}{ccc}
1&0&0\\
0&\omega^2&0\\
0&0&\omega
\end{array}
\right)
\end{array}
\end{eqnarray*}
where $\omega=e^{2\pi i/3}=(-1+\sqrt{3})/2$. In the identity
representation $1_1$, all the elements are mapped onto the number 1.
In the antisymmetric representation $1_2$, the group elements
correspond to 1 or -1 respectively for even permutation and odd
permutation. For the 2 representation, the representation matrices
are as follows
\begin{eqnarray}
\nonumber&&{\cal C}_1:\;\left(\begin{array}{cc} 1&0\\
0&1
\end{array}\right)\\
\nonumber&&{\cal C}_2:\; STS^2=T^2S=\left(\begin{array}{cc}
0&\omega^2\\
\omega&0
\end{array}\right),~~TSTS^2=TST=\left(\begin{array}{cc}
0&1\\
1&0
\end{array}\right),~~
ST^2=S^2TS=\left(\begin{array}{cc}
0&\omega\\
\omega^2&0
\end{array}\right)\\
\nonumber&&{\cal C}_3:\; TS^2T^2=S^2=T^2S^2T=\left(
\begin{array}{cc}
1&0\\
0&1
\end{array}
\right)\\
\nonumber&&{\cal C}_4:\;T=S^2T=TS^2=S^2TS^2=\left(\begin{array}{cc}\omega&0\\
0&\omega^2
\end{array}\right),~~T^2=S^2T^2=STS=T^2S^2=\left(\begin{array}{cc}\omega^2&0\\
0&\omega
\end{array}\right)\\
\nonumber&&{\cal C}_5:\; S=S^3=\left(\begin{array}{cc} 0&1\\ 1&0
\end{array}\right),~~T^2ST=TS=\left(\begin{array}{cc} 0&\omega\\ \omega^2&0
\end{array}\right),~~ST=TST^2=\left(\begin{array}{cc} 0&\omega^2\\ \omega&0
\end{array}\right)
\end{eqnarray}
For the $3_1$ representation, the representation matrices are
\begin{eqnarray*}
\nonumber&&{\cal C}_1:\;\left(\begin{array}{ccc} 1&0&0\\
0&1&0\\
0&0&1
\end{array}\right)\\
\nonumber&&{\cal C}_2:\; STS^2=\left(\begin{array}{ccc}
1&0&0\\
0&0&\omega\\
0&\omega^2&0
\end{array}\right),~~TSTS^2=\left(\begin{array}{ccc}
1&0&0\\
0&0&1\\
0&1&0
\end{array}\right),~~ST^2=\frac{1}{3}\left(\begin{array}{ccc}
-1&2\omega^2&2\omega\\
2\omega&2&-\omega^2\\
2\omega^2&-\omega&2
\end{array}\right),\\
\nonumber&&~~~~~~~S^2TS=\left(\begin{array}{ccc}
1&0&0\\
0&0&\omega^2\\
0&\omega&0
\end{array}\right),~~TST=\frac{1}{3}\left(\begin{array}{ccc}
-1&2&2\\
2&2&-1\\
2&-1&2
\end{array}\right),~~T^2S=\frac{1}{3}\left(\begin{array}{ccc}
-1&2\omega&2\omega^2\\
2\omega^2&2&-\omega\\
2\omega&-\omega^2&2
\end{array}\right)\\
\nonumber&&{\cal C}_3:\;TS^2T^2=\frac{1}{3}\left(
\begin{array}{ccc}
-1&2\omega&2\omega^2\\
2\omega^2&-1&2\omega\\
2\omega&2\omega^2&-1
\end{array}
\right),S^2=\frac{1}{3}\left(
\begin{array}{ccc}
-1&2&2\\
2&-1&2\\
2&2&-1
\end{array}
\right),T^2S^2T=\frac{1}{3}\left(
\begin{array}{ccc}
-1&2\omega^2&2\omega\\
2\omega&-1&2\omega^2\\
2\omega^2&2\omega&-1
\end{array}
\right)\\
\nonumber&&{\cal C}_4:\;T=\left(\begin{array}{ccc} 1&0&0\\
0&\omega^2&0\\
0&0&\omega
\end{array}\right),~~T^2=\left(\begin{array}{ccc} 1&0&0\\
0&\omega&0\\
0&0&\omega^2
\end{array}\right),~~T^2S^2=\frac{1}{3}\left(\begin{array}{ccc}-1&2&2\\
2\omega&-\omega&2\omega\\
2\omega^2&2\omega^2&-\omega^2\\
\end{array}\right),\\
\nonumber&&~S^2T=\frac{1}{3}\left(\begin{array}{ccc}-1&2\omega^2&2\omega\\
2&-\omega^2&2\omega\\
2&2\omega^2&-\omega
\end{array}\right),S^2TS^2=\frac{1}{3}\left(\begin{array}{ccc}-1&2\omega&2\omega^2\\
2\omega&-\omega^2&2\\
2\omega^2&2&-\omega
\end{array}\right),STS=\frac{1}{3}\left(\begin{array}{ccc}-1&2\omega^2&2\omega\\
2\omega^2&-\omega&2\\
2\omega&2&-\omega^2
\end{array}\right),\\
\nonumber&&~~S^2T^2=\frac{1}{3}\left(\begin{array}{ccc}-1&2\omega&2\omega^2\\
2&-\omega&2\omega^2\\
2&2\omega&-\omega^2
\end{array}\right),~~TS^2=\frac{1}{3}\left(\begin{array}{ccc}-1&2&2\\
2\omega^2&-\omega^2&2\omega^2\\
2\omega&2\omega&-\omega
\end{array}\right)\\
\nonumber&&{\cal C}_5:\;
S=\frac{1}{3}\left(\begin{array}{ccc} -1&2\omega&2\omega^2\\
2\omega&2\omega^2&-1\\
2\omega^2&-1&2\omega
\end{array}\right),T^2ST=\frac{1}{3}\left(\begin{array}{ccc} -1&2&2\\
2\omega^2&2\omega^2&-\omega^2\\
2\omega&-\omega&2\omega
\end{array}\right),ST=\frac{1}{3}\left(\begin{array}{ccc}-1&2&2\\
2\omega&2\omega&-\omega\\
2\omega^2&-\omega^2&2\omega^2
\end{array}\right),\\
\nonumber&&~~~~TS=\frac{1}{3}\left(\begin{array}{ccc} -1&2\omega&2\omega^2\\
2&2\omega&-\omega^2\\
2&-\omega&2\omega^2
\end{array}\right),~TST^2=\frac{1}{3}\left(\begin{array}{ccc}-1&2\omega^2&2\omega\\
2&2\omega^2&-\omega\\
2&-\omega^2&2\omega
\end{array}\right),~S^3=\frac{1}{3}\left(\begin{array}{ccc} -1&2\omega^2&2\omega\\
2\omega^2&2\omega&-1\\
2\omega&-1&2\omega^2
\end{array}\right)
\end{eqnarray*}
Since the signs of the generator $S$ are opposite in $3_1$ and $3_2$
representations, the represention matrices for the $3_2$
representation can be found from those of the $3_1$ representation:
the matrices are exactly the same for ${\cal C}_1$, ${\cal C}_3$ and
${\cal C}_4$ classes, whereas they are the opposite for ${\cal C}_2$
and ${\cal C}_5$.

\section*{Appendix B: NLO corrections to the vacuum alignment}

In this appendix, we will analyze the NLO corrections to the vacuum
alignment induced by the higher dimensional operators. At NLO the
driving superpotential dependent on the driving fields is modified
into
\begin{equation}
\label{b1}w_{v}+\Delta w_{v}
\end{equation}
where $w_v$ is the LO contributions shown in Eq.(\ref{7}), which are
of dimension three. $\Delta w_v$ is most general set of terms
suppressed by one power of the cutoff, which are linear in the
driving fields and are invariant under the symmetry of the model.
Concretely $\Delta w_v$ is given by
\begin{eqnarray}
\label{b2}\Delta w_{v}=\frac{1}{\Lambda}\Big[\sum^{2}_{i=1}p_i{\cal O}^{\varphi}_i+\sum^5_{i=1}x_i{\cal O}^{\xi'}_i+\sum^3_{i=1}t_i{\cal O}^{\theta}_i+\sum^{2}_{i=1}e_i{\cal O}^{\eta}_i+\sum^{2}_{i=1}s_i{\cal O}^{\phi}_i\Big]
\end{eqnarray}
where $p_i$, $x_i$, $t_i$, $e_i$ and $s_i$ are order one coefficients, $\{{\cal O}^{\varphi}_i,{\cal O}^{\xi'}_i,{\cal O}^{\theta}_i,{\cal O}^{\eta}_i,{\cal O}^{\phi}_i\}$ are the complete set of
invariant operators of dimension four,
\begin{eqnarray}
\label{b3}{\cal O}^{\varphi}_1=(\varphi^{0}(\varphi\chi)_{3_2})_{1_2}\,\theta,~~~~{\cal O}^{\varphi}_2=(\varphi^{0}(\eta\phi)_{3_2})_{1_2}\,\Delta
\end{eqnarray}
\begin{eqnarray}
\label{b4}{\cal O}^{\xi}_1=\xi'^{0}\theta(\varphi^2)_{1_1},~~~{\cal
O}^{\xi}_2=\xi'^{0}\theta(\chi^2)_{1_1},~~~{\cal
O}^{\xi}_3=\xi'^{0}\Delta(\eta^2)_{1_1}, ~~~{\cal
O}^{\xi}_4=\xi'^{0}\Delta(\phi^2)_{1_1},~~~{\cal
O}^{\xi}_5=\xi'^{0}\Delta^3
\end{eqnarray}
\begin{eqnarray}
\label{b5}{\cal O}^{\theta}_1=\theta^{0}(\eta^{3})_{1_1},~~~~{\cal O}^{\theta}_2=\theta^{0}(\phi^3)_{1_1},~~~~{\cal O}^{\theta}_3=\theta^{0}(\eta\phi^2)_{1_1}
\end{eqnarray}
\begin{eqnarray}
\label{b6}{\cal O}^{\eta}_1=(\eta^{0}(\eta^2)_2)_{1_2}\theta,~~~{\cal O}^{\eta}_2=(\eta^{0}(\phi^2)_{2})_{1_2}\theta
\end{eqnarray}
\begin{eqnarray}
\label{b7}{\cal O}^{\phi}_1=(\phi^{0}(\phi^2)_{3_1})_{1_2}\theta,~~~~{\cal O}^{\phi}_2=(\phi^{0}(\eta\phi)_{3_1})_{1_2}\theta
\end{eqnarray}
The NLO superpotential $\Delta w_{v}$ results in shift of the LO
VEVs, then the vacuum configuration is modified into
\begin{eqnarray}
\nonumber&&\langle\varphi\rangle=(\delta v_{\varphi_1},v_{\varphi}+\delta v_{\varphi_2},\delta v_{\varphi_3}),~~~~\langle\chi\rangle=(\delta v_{\chi_1},v_{\chi},\delta v_{\chi_3}),\\
\nonumber&&\langle\eta\rangle=(v_{\eta},v_{\eta}+\delta v_{\eta_2}),~~~~~~\langle\phi\rangle=(v_{\phi}+\delta v_{\phi_1},v_{\phi}+\delta v_{\phi_2},v_{\phi}+\delta v_{\phi_3}),\\
\label{b8}&&\langle\theta\rangle=v_{\theta}+\delta v_{\theta} 
\end{eqnarray}
Note that the VEV $v_{\Delta}$ doesn't receive correction at NLO.
Similar to section \ref{sec:alignment}, the new vacuum configuration
is obtained by imposing the vanish of the first derivative of
$w_v+\Delta w_v$ with respect to the driving fields $\varphi^{0}$,
$\xi'^{0}$, $\theta^{0}$, $\eta^{0}$ and $\phi^{0}$. Only terms
linear in the shift $\delta v$ are kept, and terms of order $\delta
v/\Lambda$ are neglected, then the minimization equations become,
\begin{eqnarray}
\nonumber&&(2g_1v_{\varphi}+g_3v_{\chi})\delta v_{\varphi_3}+(2g_2v_{\chi}-g_3v_{\varphi})\delta v_{\chi_3}=0\\
\nonumber&&2g_1\delta v_{\varphi_2}+p_1\frac{v_{\chi}v_{\theta}}{\Lambda}=0\\
\nonumber&&(2g_1v_{\varphi}-g_3v_{\chi})\delta v_{\varphi_1}+(2g_2v_{\chi}+g_3v_{\varphi})\delta v_{\chi_1}=0\\
\nonumber&&g_4(v_{\varphi}\delta v_{\chi_3}+v_{\chi}\delta v_{\varphi_3})+\frac{1}{\Lambda}(2x_3v_{\Delta}v^2_{\eta}+3x_4v_{\Delta}v^2_{\phi}+x_5v^3_{\Delta})=0\\
\label{b9}&&\kappa v_{\theta}\delta v_{\theta}+\frac{1}{\Lambda}(t_1v^3_{\eta}+3t_3v_{\eta}v^2_{\phi})=0
\end{eqnarray}
Solving the above linear equations, we obtain
\begin{eqnarray}
\nonumber&&\delta v_{\varphi_1}=\frac{2g_2v_{\chi}+g_3v_{\varphi}}{
g_3v_{\chi}-2g_1v_{\varphi}}\delta v_{\chi_1}\\
\nonumber&&\delta v_{\varphi_2}=-\frac{p_1}{2g_1}\frac{v_{\chi}v_{\theta}}{\Lambda}\\
\nonumber&&\delta v_{\varphi_3}=\frac{2g_2v_{\chi}-g_{3}v_{\varphi}}{2g_4\Lambda(2g_1v^2_{\varphi}+g_3v_{\chi}v_{\varphi})}[2x_3v_{\Delta}v^2_{\eta}+3x_4v_{\Delta}v^2_{\phi}+x_5v^3_{\Delta}]\\
\nonumber&&\delta v_{\chi_3}=-\frac{2g_1v_{\varphi}+g_3v_{\chi}}{2g_4\Lambda(2g_1v^2_{\varphi}+g_3v_{\chi}v_{\varphi})}[2x_3v_{\Delta}v^2_{\eta}+3x_4v_{\Delta}v^2_{\phi}+x_5v^3_{\Delta}]\\
\label{b10}&&\delta v_{\theta}=-\frac{t_1}{\kappa}\frac{v^3_{\eta}}{\Lambda v_{\theta}}-\frac{3t_3}{\kappa}\frac{v_{\eta}v^2_{\phi}}{\Lambda v_{\theta}}
\end{eqnarray}
where $\delta v_{\varphi_{2,3}}$, $\delta v_{\chi_3}$ and $\delta
v_{\theta}$ are of order $1/\Lambda$, $\delta v_{\chi_{1}}$ is
undetermined, and it is expected to be suppressed by $1/\Lambda$ as
well. The minimization conditions for the shift $\delta v_{\eta_2}$
and $\delta v_{\phi_{1,2,3}}$ are
\begin{eqnarray}
\nonumber&&2f_2v_{\phi}(\delta v_{\phi_1}+\delta v_{\phi_2}+\delta v_{\phi_3})+e_1\frac{v^2_{\eta}v_{\theta}}{\Lambda}+3e_2\frac{v^2_{\phi}v_{\theta}}{\Lambda}=0\\
\nonumber&&2f_1v_{\eta}\delta v_{\eta_2}+2f_2v_{\phi}(\delta v_{\phi_1}+\delta v_{\phi_2}+\delta v_{\phi_3})-e_1\frac{v^2_{\eta}v_{\theta}}{\Lambda}-3e_2\frac{v^2_{\phi}v_{\theta}}{\Lambda}=0\\
\nonumber&&f_{3}(v_{\eta}\delta v_{\phi_2}-v_{\eta}\delta v_{\phi_3}-v_{\phi}\delta v_{\eta_2})+2s_2\frac{v_{\eta}v_{\phi}v_{\theta}}{\Lambda}=0\\
\nonumber&&f_{3}(v_{\eta}\delta v_{\phi_1}-v_{\eta}\delta v_{\phi_2}-v_{\phi}\delta v_{\eta_2})+2s_2\frac{v_{\eta}v_{\phi}v_{\theta}}{\Lambda}=0\\
\label{b11}&&f_{3}(v_{\eta}\delta v_{\phi_3}-v_{\eta}\delta
v_{\phi_1}-v_{\phi}\delta
v_{\eta_2})+2s_2\frac{v_{\eta}v_{\phi}v_{\theta}}{\Lambda}=0
\end{eqnarray}
The solutions to the above equations are
\begin{eqnarray}
\nonumber&&\delta v_{\eta_2}=\frac{2s_2}{f_3}\frac{v_{\eta}v_{\theta}}{\Lambda}\\
\label{b12}&&\delta v_{\phi}\equiv\delta v_{\phi_1}=\delta v_{\phi_2}=\delta v_{\phi_3}=-\frac{e_1}{6f_2}\frac{v^2_{\eta}v_{\theta}}{\Lambda v_{\phi}}-\frac{e_2}{2f_2}\frac{v_{\phi}v_{\theta}}{\Lambda}
\end{eqnarray}
We notice that $\langle\phi\rangle$ acquires ${\cal O}(1/\Lambda)$ corrections in the same directions, and the shift $\delta v_{\eta_2}$ is in general non-zero. Since
all the VEVs approximately are of the same order ${\cal O}(\lambda^2_c\Lambda)$ at LO, we expect $\frac{\delta VEV}{VEV}\sim\frac{VEV}{\Lambda}\sim\lambda^2_c$. Therefore the LO vacuum alignment in Eq.(\ref{9}) and Eq.(\ref{12}) is stable under the NLO corrections.

\end{document}